\documentclass[aps,rmp,superscriptaddress,amsmath,amssymb,floatfix,
,reprint
]{revtex4-1}
\usepackage[english]{babel}

\usepackage{newtxtext}
\usepackage[nosymbolsc,bigdelims,varg]{newtxmath}
\DeclareMathAlphabet{\mathcal}{OMS}{cmsy}{m}{n}
\DeclareSymbolFont{largesymbolsCM}{OMX}{cmex}{m}{n}
\let\sum\relax
\DeclareMathSymbol{\sum}{\mathop}{largesymbolsCM}{"50}

\usepackage{graphicx}

\usepackage{color}
\definecolor{darkblue}{rgb}{0,0,.6}
\definecolor{darkgreen}{rgb}{0,0.5,0}

\usepackage{ifpdf,tabularx}

\ifpdf
  \usepackage{epstopdf}
  \usepackage[pdftex,unicode,
  pdfstartview={FitH},pdfborder={0 0 0}]{hyperref}
  \usepackage{hypcap}
\else
  \usepackage[hypertex]{hyperref}
\fi
\hypersetup{
  bookmarksnumbered = true,
  colorlinks = true, linkcolor = darkblue,
  citecolor = darkblue, filecolor = darkblue,
  menucolor = darkblue, urlcolor = darkblue
}

\usepackage{bm}
\usepackage{enumitem}
\setlist{nosep}
\usepackage{microtype}

\let\vec=\bm
\newcommand{\abs}[1]{\ensuremath{\left| #1 \right|}}
\let\Re\undefined
\DeclareMathOperator{\Re}{Re}
\let\Im\undefined
\DeclareMathOperator{\Im}{Im}
\DeclareMathOperator{\Ai}{Ai}

\newcommand{\e}{\mathrm{e}}
\newcommand{\w}[1]{\omega_{\mathrm{#1}}}

\newcommand{\tw}[1]{\widetilde{\omega}_\mathrm{#1}}
\newcommand{\tW}[1]{\widetilde{\Omega}_\mathrm{#1}}
\newcommand{\g}[1]{\gamma_{\mathrm{#1}}}

\newcommand{\E}[1]{E_{\mathrm{#1}}}
\newcommand{\ii}{\mathrm{i}}
\newcommand{\dd}{\mathrm{d}}
\newcommand{\Up}{U_\mathrm{p}}
\newcommand{\ellmax}{\ell_\mathrm{max}}
\newcommand{\um}{\textmu{}m}
\newcommand{\brK}[1]{\ensuremath{\bm{(}\vec K(#1)\bm{)}}}
\newcommand{\brKx}[1]{\ensuremath{\bm{(}K_x(#1)\bm{)}}}
\newcommand{\brL}[1]{\ensuremath{\bm{(}\vec\Lambda(#1)\bm{)}}}

\newcolumntype{C}{>{$\displaystyle}c<{$}}
\begin{document}

\title{\textit{Colloquium}: Strong-field Phenomena in Periodic Systems}

\author{Stanislav Yu.~Kruchinin}
\email{stanislav.kruchinin@univie.ac.at}
\affiliation{Max-Planck-Institut f\"ur Quantenoptik, Hans-Kopfermann-Str. 1, 85748 Garching, Germany}
\affiliation{University of Vienna, Faculty of Physics and Center for Computational Materials Sciences, Sensengasse 8/12, 1090 Vienna, Austria}

\author{Ferenc Krausz}
\email{ferenc.krausz@mpq.mpg.de}
\affiliation{Max-Planck-Institut f\"ur Quantenoptik, Hans-Kopfermann-Str. 1, 85748 Garching, Germany}
\affiliation{Ludwig-Maximilians-Universit\"at, Am Coulombwall~1, 85748 Garching, Germany}

\author{Vladislav S.~Yakovlev}
\email{vladislav.yakovlev@mpq.mpg.de}
\affiliation{Max-Planck-Institut f\"ur Quantenoptik, Hans-Kopfermann-Str. 1, 85748 Garching, Germany}
\affiliation{Ludwig-Maximilians-Universit\"at, Am Coulombwall~1, 85748 Garching, Germany}

\date{\today}

\begin{abstract}
The advent of visible-infrared laser pulses carrying a substantial fraction of their energy in a single field oscillation cycle has opened a new era in the experimental investigation of ultrafast processes in semiconductors and dielectrics (bulk as well as nanostructured), motivated by the quest for the ultimate frontiers of electron-based signal metrology and processing.
Exploring ways to approach those frontiers requires insight into the physics underlying the interaction of strong high-frequency (optical) fields with electrons moving in periodic potentials.
This Colloquium aims at providing this insight.
Introduction to the foundations of strong-field phenomena defines and compares regimes of field--matter interaction in periodic systems, including (perfect) crystals as well as optical and semiconductor superlattices, followed by a review of recent experimental advances in the study of strong-field dynamics in crystals and nanostructures.
Avenues toward measuring and controlling electronic processes up to petahertz frequencies are discussed.
\end{abstract}


\maketitle

\tableofcontents{}

\section{Introduction}

The description of a particle in a periodic potential has been one of the greatest successes of quantum mechanics, revolutionizing solid-state physics and triggering the evolution of modern electronics, which forms the basis of our life.
First works on band theory~\cite{Bloch_1929_ZP_52_555, Bethe_1928_AP_392_55, Kronig_1931_PRSLA_130_499}, localized electronic states~\cite{Wannier_1960_PR_117_432, Wannier_1962_RMP_34_645}, and electron dynamics in the presence of an external field~\cite{Zener_1934_PRSA_145_523, Houston_1940_PR_57_184, Kane_1960_JPCS_12_181, Keldysh_1965_JETP_20_1307} contributed to the foundations of strong-field semiconductors physics and its numerous industrial applications.

Progress in studies of optical and transport phenomena in solids governed the evolution of modern electronics.
While first-generation semiconductor devices were based on a diffusive transport, today's nano-MOSFETs (metal-oxide-semiconductor field-effect transistor) operate in the quasiballistic and high-field regimes~\cite{Frank_2001_PI_89_259, Martin_2004_ITED_51_1148, Palestri_2005_ITED_52_2727, Datta_2012}.
As a result of the continued miniaturization of semiconductor integrated circuits~\cite{Taur_2013, Ionescu_2011_Nature_479_329}, microelectronics has approached operating regimes where dissipation phenomena limit the rate of information processing~\cite{Pop_2010_NR_3_147, Markov_2014_Nature_512_147}.
For example, in spite of ballistic transport within nm-scale semiconductor channels, the clock frequency in contemporary digital electronics cannot be increased beyond several gigahertz because of excessive energy dissipation in the contacts~\cite{Pop_2010_NR_3_147, Datta_2012}.

Recent developments in the synthesis of intense light pulses with a precisely controlled electric field in the terahertz~\cite{Liu_1996_IQE_2_709, Ferguson_2002_NM_1_26, Kohler_2002_Nature_417_156, Chan_2011_Sci_331_6021}, infrared, and visible domains~\cite{Baltuska_2003_Nature_421_611, Goulielmakis_2008_Sci_320_1614, Huang_2011_NP_5_475, Fattahi_2014_Optica_1_45} present a new approach to understanding and, possibly, overcoming the speed limits of ultrafast solid-state metrology and spectroscopy~\cite{Agostini_2004_RPP_67_813, Krausz_2009_RMP_81_163, Krausz_2014_NP_8_205}.
Strong light fields are able to substantially and nondestructively modify electronic and optical properties of a solid within a single oscillation of the field~\cite{Schiffrin_2013_Nature_493_70, Schultze_2013_Nature_493_75, Schubert_2014_NP_8_119, Lucchini_2016_Science_353_916}.
The relevant effects may last only as long as the external field is present, as in the case of the dynamic Franz--Keldysh effect, or they may be followed by relatively slow relaxation dynamics, as in the case of interband excitations.
Strong fields also enable novel applications, such as the high-harmonic generation in bulk solids~\cite{Ghimire_2011_NP_7_138, Schubert_2014_NP_8_119, Luu_2015_Nature_521_498, Vampa_2014_PRL_113_073901, Ndabashimiye_2016_Nature_534_520, Hammond_2017_NP_11_594}, two-dimensional (2D) materials~\cite{Al-Naib_2014_PRB_90_245423, Liu_2017_NP_13_262, Cox_2017_NC_8_14380},
and artificially designed plasmonic structures~\cite{Han_2016_NC_7_13105, Vampa_2017_NP_13_659, Sivis_2017_Sci_357_303, Ciappina_2017_RPP_80_054401}.

Attosecond metrology offers experimental means for studying and steering processes that unfold within a cycle of an optical field, promising direct access to light-controlled electron motion in a regime where quantum coherence is preserved.
Nevertheless, pushing the frontiers of solid-state metrology to multi-PHz bandwidths and sub-100-attosecond temporal resolution is a major challenge that requires further technological progress and new insight into the basic physics of light-driven electron motion.
In this context, valuable lessons can be learned from research on artificial periodic structures, such as semiconductor superlattices~\cite{Tsu_2011, Ivchenko_1997, Leo_2003} and optical lattices~\cite{Bloch_2005_NP_1_23, Lewenstein_2012, Gardiner_2014}, where similar physical processes take place under conditions more convenient for experiments.
Reviewing the basic physical phenomena that may affect the future evolution of ultrafast metrology and signal processing, we combine insight from several disparate scientific communities.

This Colloquium is structured as follows.
This Introduction is followed by a classification of field--matter interaction regimes in terms of dimensionless parameters, where we also recapitulate the key approximations for simplified modeling of ultrafast phenomena in periodic systems (Sec.~\ref{s:2}).
In Sec.~\ref{s:3}, we review widely known strong-field phenomena in periodic structures.
Discussing similarities and differences between natural and artificial systems, we focus on bulk solids and semiconductor superlattices.
We address new measurement and control techniques offered by attosecond science, such as field-resolved control of transport properties at optical frequencies, high-order harmonic radiation spectroscopy, and real-time probing of the electronic processes unfolding under the influence of strong, controlled optical fields.
Finally, we conclude with an outlook into the future of attosecond solid-state physics (Sec.~\ref{s:4}).

\section{Interaction regimes}\label{s:2}

\subsection{Basic concepts}

In our Colloquium, we discuss a range of light--matter interaction regimes that involve different theoretical approaches and physical interpretations.
Searching for a unified description of these regimes, we classify them using a set of dimensionless parameters, each of which is a ratio of two characteristic frequencies.
They describe the incident light, the solid, and the most basic physical phenomena observed in the solid interacting with light.
We frequently refer to these dimensionless quantities as \emph{adiabaticity parameters}~\cite{Mostafazadeh_1997_PRA_55_1653}.
For example, the ratio of the fundamental band gap $\E{g}$ to a photon energy $\hbar\w0$ is one such parameter.
It quantifies the degree of adiabaticity because the characteristic response time of valence electrons to an external perturbation is inversely proportional to the band gap.
So, if $N = \E{g} / \hbar\w0 \gg 1$, then the wavefunction of a valence electron adiabatically adapts itself to the external field unless the field is strong enough to let the electron escape the binding potential.
The concept of adiabaticity has proven to be a valuable analysis tool, giving insight into the dynamics of complex systems whenever very different and thus well-separable temporal or spatial scales are involved.

For an electron driven by a homogeneous electric field $\vec F(t)$ in a spatially periodic lattice potential, intraband motion constitutes an important example of adiabatic dynamics.
In accord with its name, this is the motion of an electron within a particular band.
If, by virtue of approximations, the electron is not allowed to undergo transitions to other bands, then the only degree of freedom left for carrier dynamics is the intraband motion, which obeys the following equations~\cite{Bloch_1929_ZP_52_555}:
\begin{subequations}\label{e:Bloch}
\begin{align}
  \label{e:Bloch_k}
  \dot{\vec K} &= -\frac{e}{\hbar} \vec F(t)
,\\
  \label{e:Bloch_r}
  \dot{\vec r}_n &= \frac{1}{\hbar} \nabla_{\vec K} E_n(\vec K)
,
\end{align}
\end{subequations}
where $e > 0$ is the elementary charge, $n$ is a band index, and $E_n(\vec k)$ is the dependence of charge-carrier energy on a crystal momentum, which is referred to as band dispersion.

The first equation~\eqref{e:Bloch_k} is known as the acceleration theorem.
It resembles the classical Newton's equation $\dot{\vec P} = -e \hbar^{-1} \vec F(t)$, describing the kinetic momentum of a free electron in an external electric field.
Nevertheless, Eq.~\eqref{e:Bloch_k} is an inherently quantum-mechanical result~\cite{Rossi_2002_RMP_74_895}.
The general solution of this equation is
\begin{equation}\label{e:Kt}
  \vec K(t) = \vec K(t_0) - \frac{e}{\hbar} \int_{t_0}^t \vec F(t_1)\,\dd t_1.
\end{equation}
This result is equally applicable if the carrier's initial state is given by a pure Bloch state or by a wavepacket with a central crystal momentum $\vec K(t_0) = \vec k$~\cite{Ashcroft_1976}.
By contrast, the second equation of motion~\eqref{e:Bloch_r} is valid only for a spatially localized wavepacket with well-defined center-of-mass coordinate $\vec r_n$ and group velocity $\vec v_n\brK{t} \equiv \dot{\vec r}_n$.

Within the approximation of a purely intraband motion, an oscillating electric field periodically changes the electron's energy.
The cycle-averaged energy of intraband motion is known as the \emph{ponderomotive energy}.
For a monochromatic electric field with frequency $\omega_0$, we consider a reciprocal-space trajectory $\vec K(t)$ oscillating around crystal momentum $\vec k$.
Averaging the $n$th band energy $E_n$ over a period of $T_0 = 2\pi / \omega_0$, we obtain
\begin{multline}\label{e:Upn}
  \Up^{(n)}(\vec k) = \frac{1}{T_0} \int_{0}^{T_0} [E_n\brK{t} - E_n(\vec k)]\,\dd t =
\\
  \overline{E_n\brK{t}} -  E_n(\vec k).
\end{multline}

As we will show in Sec.~\ref{s:Keldysh}, the probabilities of transitions between two energy bands depend on the effective band gap~$\E{cv}\brK{t} = \E{c}\brK{t} - \E{v}\brK{t}$, which is influenced by the intraband motion.
To evaluate the cycle-averaged band gap, one needs to subtract the ponderomotive energy in the initial state from that in the final state: $\Up(\vec k) = \Up^{(\mathrm{c})}(\vec k) - \Up^{(\mathrm{v})}(\vec k)$, where $n = \mathrm{c}$ and $n = \mathrm{v}$ denote the lowest conduction and highest valence band, respectively.
This difference yields the ponderomotive energy of an electron-hole pair:
\begin{equation}\label{e:Up}
  \Up(\vec k) = \overline{\E{cv}\brK{t}} - \E{cv}(\vec k),
\end{equation}
where $\E{cv}(\vec k)$ is the band gap at crystal momentum $\vec k$.
In direct-band-gap materials, $ \E{cv}(0) = \E{g}$.

We use the term ``electron-hole pair'' because on the time scales much shorter than those of momentum scattering and dephasing emerging carriers are described by a coherent superposition of states in the valence and conduction bands and thus cannot be considered as independent particles.
Equation~\eqref{e:Kt} applies equally to a conduction-band (CB) electron and the hole left in the valence band, hence their intraband motion after excitation is correlated, even though the external field tends to separate them in real space.

In the effective-mass approximation (EMA), the band energies are given by~\cite{Yu_2010}
\begin{equation}\label{e:EnkEMA}
  E_n^\mathrm{(EMA)}(\vec k) = \frac{\hbar^2\vec k^2}{2 m_n},
\end{equation}
where, for simplicity, we assumed an isotropic medium.
Using Eq.~\eqref{e:EnkEMA} we obtain the following expression for the ponderomotive energy of an electron-hole pair:
\begin{equation}\label{e:UpEMA}
  \Up^\mathrm{(EMA)} = \frac{e^2 F_0^2(1 + \beta^2)}{4m\omega_0^2}
,
\end{equation}
where $F_0$ is the amplitude of the oscillating electric field,
\begin{equation*}
  \beta = \frac{|F_+| - |F_-|}{|F_+| + |F_-|}
\end{equation*}
is the ellipticity, and $F_+$ and $F_-$ are the amplitudes of left- and right-rotating components, respectively.
In particular, $\beta = 0$ stands for the linear polarization and $\beta = \pm 1$ corresponds to the circular one.
Equation~\eqref{e:UpEMA} involves the reduced effective mass of an electron-hole pair:
\begin{equation}\label{e:m}
  \frac{1}{m} = \frac{1}{m_\mathrm{c}} - \frac{1}{m_\mathrm{v}}
              = \frac{1}{m_\mathrm{c}} + \frac{1}{m_\mathrm{h}},
\end{equation}
where $m_\mathrm{h} = -m_\mathrm{v}$ is the hole mass.
Ponderomotive energy is also well defined for nonparabolic bands.
In Appendix~\ref{s:Up}, we provided an analytical expression for $\Up$ in the tight-binding approximation.

An important characteristic frequency that we used in our classification scheme is inversely proportional to the time it takes $\vec K(t)$ to make a round-trip through the first Brillouin zone (BZ) in a static field $\vec F_0$.
For any reciprocal lattice vector $\vec G$, crystal momenta $\vec k$ and $\vec k + \vec G$ are equivalent, so in the absence of interband transitions, the wavefunction of an electron exposed to a static external field parallel to $\vec G$ would oscillate with a frequency known as the \emph{Bloch frequency}:
\begin{equation}
  \w{B} = \frac{2\pi |e F_0|}{\hbar G} = \frac{|e F_0| a}{\hbar},
\end{equation}
where $a$ is the lattice constant.

Finally, we also used the peak Rabi frequency
\begin{equation}\label{e:wR}
  \w{R} = \frac{|e \vec F_0\cdot\vec\xi_\mathrm{cv}^\mathrm{(max)}|}{\hbar}
\end{equation}
to classify strong-field phenomena occurring in a nearly resonant field with amplitude $\vec F_0$.
Here $|\vec\xi_\mathrm{cv}^\mathrm{(max)}| = \max_{\vec k} |\vec\xi_\mathrm{cv}(\vec k)|$ is the peak absolute value of the interband matrix element, which we defined later by Eq.~\eqref{e:xinn}.

Considering the ratios of characteristic frequencies, we obtain a set of dimensionless parameters summarized in Table~\ref{t:gamma} that describe various regimes of interaction between an optical field and a solid.
The exact definitions and physical interpretations of these parameters are given in the rest of this section where we discuss physical processes inherent to different regimes.

\renewcommand{\arraystretch}{1.5}
\begin{table}[t]
  \caption{\label{t:gamma}%
    Characteristic frequencies and dimensionless parameters describing the regimes (adiabatic or diabatic limits) of laser field interaction with periodic potentials.
    The parameters are proportional to the ratio of the frequencies from the upper row and the left column.
  }
  \begin{tabular}{C|CCCC}
    & \E{g}/\hbar & \Up/\hbar & \w{B}      & \w{R} \\ \hline
    \w{0}       & N           & \g{NP}     & \g{DL} & \g{RF}^{(0)}        \\
    \E{g}/\hbar &             & \g{K}^{-2} & \g{BH} & \g{RF}^\mathrm{(g)} \\
    \Up/\hbar   &             &            & \g{BP} & \g{RP}              \\
    \w{B}       &             &            &        & \g{RB}              \\
  \end{tabular}
\end{table}

Since the concept of adiabaticity plays an important role for strong-field phenomena, we briefly review the corresponding quantum-mechanical formalism.
Let the Hamilton operator be $\hat H(t) = \hat H_0 + \hat V(t)$, where $\hat H_0$ is the Hamiltonian of the unperturbed quantum system, and $\hat V(t)$ accounts for the interaction with an external time-dependent field that is present only during a finite interval of time.
The interaction is not necessarily weak, and we describe it with the time-dependent Schr\"odinger equation (TDSE):
\begin{equation}\label{e:TDSE}
  i\hbar\frac{\dd \Psi(\vec r, t)}{\dd t} = \hat H(t) \Psi(\vec r, t).
\end{equation}

If the system is initially in an eigenstate of $\hat H_0$, then the adiabatic theorem of quantum mechanics~\cite{Born_1928_ZP_51_165} dictates that the system must remain in an adiabatic (instantaneous) eigenstate of the time-dependent Hamiltonian as long as the interaction potential $\hat V(t)$ changes sufficiently slowly and the state is nondegenerate.
Frequently, the Hamiltonian depends on the time via a set of adiabatic parameters $\vec\Lambda(t)$: $\hat H(t) \equiv \hat H\brL{t}$.
Then the adiabatic states are defined by the following eigenvalue problem:
\begin{equation}
  \hat{H}\brL{t} \psi_{n}\brL{t} = E_{n}\brL{t} \psi_{n}\brL{t}.
\end{equation}
For the simplest case of discrete and nondegenerate levels, this approximation is well justified if the potential $\hat V(t)$ changes on a time scale much longer than the characteristic time $\tau = 2\pi\hbar/|E_n\brL{t} - E_m\brL{t}|$~\cite{Sakurai_2013}:
\begin{equation*}
  \tau \abs{\frac{\partial \hat V(t)}{\partial t}} \ll \abs{\hat V(t)}.
\end{equation*}
Also, there are modern generalizations of the adiabatic theorem that allow for degeneracies~\cite{Avron_1999_CMP_203_445, Rigolin_2012_PRA_85_062111}.

A rapidly varying $\hat V(t)$ can prevent the wavefunction from adapting to it.
Hence the system will undergo transitions between adiabatic states.
This type of evolution is called \emph{diabatic} or, equivalently, \emph{nonadiabatic}.
A general evolution of a quantum system can be described by a superposition of adiabatic states:
\begin{equation}\label{e:Psirt}
  \Psi(\vec r, t) = \sum_n a_n(t) \psi_n\brL{t}.
\end{equation}
Substituting Eq.~\eqref{e:Psirt} into the TDSE~\eqref{e:TDSE} yields equations for the probability amplitudes $a_n(t)$.
This procedure provides a general framework for iterative solution of the TDSE in the basis of adiabatic eigenstates known as \emph{adiabatic perturbation theory}~\cite{Bransden_2000, Teufel_2003, Rigolin_2008_PRA_78_052508, Sakurai_2013}.

For a specific problem, one can arrive at various adiabatic bases by taking advantage of the gauge freedom, that is, by applying unitary transformations to the Hamiltonian.
Such a transformation consists of replacing $\Psi$ with $\hat U \Psi$ and $\hat H(t)$ with $\hat U \hat H \hat U^\dagger - i \hbar \hat U (\partial\hat U^\dagger/\partial t)$, where $\hat U$ is a unitary operator.

Let us now consider a single electron moving in a periodic lattice potential, the field-free eigenstates of which are given by Bloch functions $\psi_{n, \vec k}^\mathrm{(B)}(\vec r) = u_{n, \vec k}(\vec r) \e^{\ii\vec k\cdot\vec r}$.
We begin by writing the time-dependent Hamiltonian in the velocity gauge and the dipole approximation:
\begin{equation*}
  \hat{H}(t) = \frac{\left[\hat{\vec p} + e \vec{A}(t) \right]^2}{2 m_0} + \hat V_\mathrm{latt}(\vec r),
\end{equation*}
where $m_0$ is the free-electron mass, $\hat{\vec p}$ is the momentum operator,
\begin{equation}\label{e:A}
  \vec A(t) = -\int_{t_0}^{t} \vec F(t')\,\dd t'
\end{equation}
is the vector potential,
and $\hat V_\mathrm{latt}(\vec r)$ is an effective periodic lattice potential created by ions and other electrons.
For simplicity, we assume the lattice potential to be static and local.
By using $\hat U_1 = \e^{-\ii \vec{k}\cdot\vec{r}}$ as a unitary transformation, we arrive at
\begin{equation}\label{e:HtKt}
  \hat{H}(t) = \frac{\left[\hat{\vec p} + \hbar \vec K(t) \right]^2}{2 m_0} + \hat V_\mathrm{latt}(\vec r).
\end{equation}
The transformed eigenstates of Eq.~\eqref{e:HtKt} are nothing else but periodic parts of the Bloch functions $u_{n,\vec K(t)}(\vec r)$, where the field-free crystal momentum $\vec k$ is replaced by the kinetic one $\vec K(t)$.

This form of the Hamiltonian, where $\vec K(t)$ is an adiabatic parameter that conforms to the acceleration theorem, is a starting point for applying the adiabatic perturbation theory to electrons in periodic potentials~\cite{Bychkov_1970_JETP_31_928, Zak_1989_PRL_62_2747, Xiao_2010_RMP_82_1959}.
With the transformation $\hat U_2 = \e^{\ii\vec K(t)\cdot\vec r}$, one obtains the adiabatic solutions
\begin{equation}\label{e:Houston}
  \psi_{n, \vec k}^\mathrm{(H)}(\vec r, t) = u_{n, \vec K(t)}(\vec r)
    \e^{\ii\vec K(t)\cdot\vec r}
    \e^{\ii\phi_{n,\vec k}(t, t_0)}
    \e^{\ii\gamma_{n,\vec k}(t, t_0)}
\end{equation}
known as Houston states in the length gauge~\cite{Houston_1940_PR_57_184, Berry_1984_PRSA_392_45, Zak_1989_PRL_62_2747}.
Here,
\begin{equation*}
  \phi_{n,\vec k}(t, t_0) = -\frac{1}{\hbar} \int_{t_0}^t E_n\brK{t_1}\,\dd t_1
\end{equation*}
is the dynamic phase,
\begin{equation}\label{e:gammank}
  \gamma_{n,\vec k}(t, t_0) = -\frac{e}{\hbar} \int_{t_0}^t \vec F(t_1) \cdot \vec \xi_{nn}\brK{t_1}\,\dd t_1
\end{equation}
is the geometric phase~\cite{Pancharatnam_1956_PIASA_44_247, Berry_1984_PRSA_392_45}, $E_n(\vec k)$ is the energy of the $n$th band, and
\begin{equation}\label{e:xinn}
  \vec \xi_{nm}(\vec k) = \frac{\ii}{v}\int_{v} u_{n,\vec k}^*(\vec r) \nabla_{\vec k} u_{m,\vec k}(\vec r)\,\dd^3 r
\end{equation}
is the matrix element of the crystal-coordinate operator evaluated by integration over the volume $v$ of a unit cell.
A diagonal element of this matrix $\vec\xi_{nn}(\vec k)$ is the Berry connection of the $n$th band.
We discussed effects related to the Berry connection in Section~\ref{s:Berry}.

\subsection{Interband transitions in strong nonresonant fields}
\label{s:Keldysh}

Transparent media are particularly well suited for studying highly nonlinear nondestructive phenomena because most of the energy of a laser pulse escapes the medium.
In dielectrics and semiconductors that are transparent within the bandwidth of a laser pulse, single-photon transitions are prohibited by the band gap.
Consequently, a valence-band electron can be excited to a conduction band only if several photons are absorbed at the same time.

For moderately intense light, these transitions may be well described by the conventional perturbation theory constructed in the basis of the unperturbed Hamiltonian eigenstates.
In this approach, corrections to the wavefunction are found as terms of the Taylor expansion with respect to electric-field amplitude $F_0$.
This series converges if the matrix element of a perturbation is smaller than the energy difference between unperturbed states: $|V_{nm}(t)| < |E_n^{(0)} - E_m^{(0)}|$~\cite{Bransden_2000}.
This condition defines the perturbative regime.
At sufficiently high intensities, the perturbative expansion with respect to the field amplitude fails (e.g., it may diverge or give qualitatively wrong predictions), in which case we refer to the regime as ``nonperturbative''.
We note, however, that this term restricts only the applicability of the conventional perturbation theory.
Corrections in the adiabatic perturbation theory may include nonanalytical smooth functions of the field amplitude, e.g., $\exp(-\alpha/|F_0|)$, describing nonperturbative phenomena.

An example of such a theory is the analytical approach to strong-field ionization of atoms and interband transitions in solids developed by Keldysh in his seminal paper~\cite{Keldysh_1965_JETP_20_1307}, which was followed by many related works~\cite{Bychkov_1970_JETP_31_928, Kovarskii_1971_PSSB_45_47, Faisal_1973_JPB_6_89, Reiss_1980_PRA_22_1786, Minasian_1986_PRB_34_963, Mishima_2002_PRA_66_033401, Gruzdev_2007_PRB_75_205106, Vanne_2007_PRA_75_033403, Hawkins_2013_PRA_87_063842, Shcheblanov_2017_PRA_96_063410}.
The key approximation of this theory is the neglect of the Coulomb interaction between an electron excited to the conduction band and the positively charged hole left behind in the valence band.
In atomic physics, this has become known as the strong-field approximation (SFA)~\cite{Reiss_1980_PRA_22_1786}.
For solids, the SFA translates into the neglect of excitonic effects.
This is a good approximation if the exciton binding energy $\E{ex}$ is much smaller than the work that the external field does on an electron over an exciton Bohr radius $a_\mathrm{B}$, i.e., $\E{ex} \ll |e F_0| a_\mathrm{B}$.
For example, in silicon, where $a_\mathrm{B} \approx 4.5$~nm and $\E{ex} \approx 15$~meV, the SFA is well justifiable for field amplitudes $|F_0| \gg 3.3$~V/\um.

The main result of the Keldysh theory is an analytical expression for the cycle-averaged rate of the interband transitions in a monochromatic electric field of arbitrary strength.
This expression, which we do not reproduce here, contains an important dimensionless parameter known as the \emph{Keldysh parameter}:
\begin{equation}\label{e:gK}
  \g{K} = \sqrt\frac{\E{g}}{4\Up}.
\end{equation}
In the effective-mass approximation [see Eq.~\eqref{e:UpEMA}], it can be written as
\begin{equation*}
  \g{K} = \frac{\w0}{|e F_0|} \sqrt{\frac{m \E{g}}{1 + \beta^2}},
\end{equation*}
where $m$ is the reduced effective mass defined by Eq.~\eqref{e:m}.

\begin{figure}[t]
  \includegraphics[width=\linewidth]{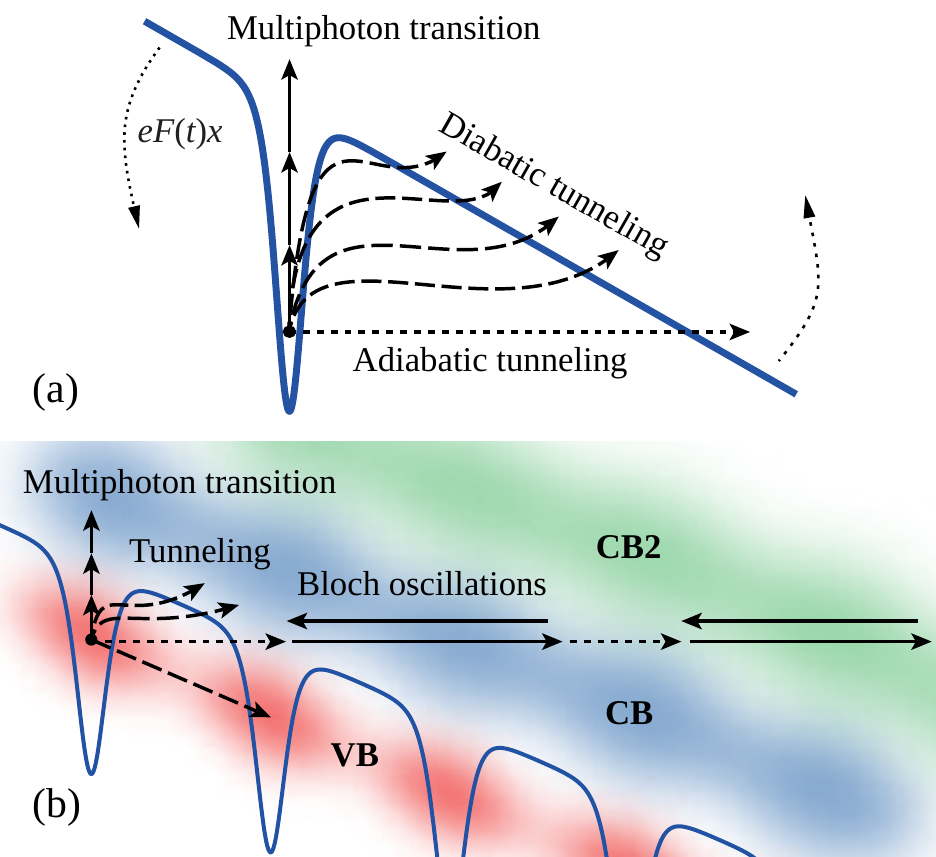}
  \caption{\label{f:potentials}%
    (Color online) Schematic real-space representation of various regimes of (a) atomic photoionization and (b) interband and intraband transitions in periodic potential exposed to an oscillating electric field.
    Here, VB, CB and CB2 denote the highest valence band, and the first and second conduction bands, respectively.
    In the multiphoton regime ($\g{K} \gg 1$), electrons are predominantly excited via absorption of energy from multiple field quanta (solid arrows).
    In the diabatic tunneling regime ($\g{K} \sim 1$), the interaction with an oscillating potential barrier transfers energy to a moving electron (long-dashed arrows).
    In the regime of adiabatic tunneling ($\g{K} \ll 1$), an electron tunnels through the potential barrier without changing its energy (horizontal short-dashed arrows).
  }
\end{figure}

This parameter classifies the regimes of interband transitions into adiabatic tunneling for $\g{K} \ll 1$, diabatic tunneling for $\g{K} \sim 1$~\cite{Nakamura_1992_JCP_97_256, Yudin_2001_PRA_64_013409, Ivanov_2005_JMO_52_165}, and multiphoton excitations for $\g{K} \gg 1$.
Figure~\ref{f:potentials} illustrates these regimes for atomic and lattice potentials in real-space representation.
In the adiabatic tunneling regime, the external field changes so slowly that the wavefunctions of bound electrons have sufficient time to adjust themselves to the evolving potentials.
The electron penetrates the partially suppressed potential barrier ``horizontally'', i.e., without changing its total energy.
In this regime, transition probabilities are well estimated by integrating static-field transition rates over time, which has been referred to as the quasistatic or adiabatic tunneling approximation.
By contrast, transitions in the multiphoton regime occur ``vertically'', without any penetration into classically forbidden regions; see Fig. 1.
These two opposite limits smoothly pass into each other in the intermediate regime known in the literature as diabatic tunneling, where the external potential changes too fast for the wavefunction of the electron to adjust, while the classically forbidden region is still involved in the process.
As a result, an electron gradually acquires some energy that facilitates its transition into the continuum of free states.

Keldysh obtained his results in the two-band approximation, assuming that the electric field predominantly excites electrons from the highest valence band to the lowest conduction band.
Substitution of Eqs.~\eqref{e:Psirt} and~\eqref{e:Houston} into the TDSE~\eqref{e:TDSE} yields the following expression for the occupation probability amplitude of the conduction-band Houston state in the two-band model~\cite{Bychkov_1970_JETP_31_928, Krieger_1986_PRB_33_5494}:
\begin{equation}\label{e:ackt}
  a_{\mathrm{c}, \vec k}^{(1)}(t) = -\frac{\ii}{\hbar} \int_{t_0}^{t} V_\mathrm{cv}(t_1)\,\e^{\ii \phi'_{\mathrm{cv}}(\vec k, t_1, t_0)}\,\dd t_1,
\end{equation}
where
\begin{equation}\label{e:wRt}
  V_\mathrm{cv}(t) = e \vec F(t)\cdot \vec \xi_\mathrm{cv}\brK{t},
\end{equation}
is the nondiagonal matrix element of interaction with the field,
\begin{equation}\label{e:phicv}
  \phi'_{\mathrm{cv}}(\vec k, t_1, t_0) = \frac{1}{\hbar} \int_{t_0}^{t_1} E'_\mathrm{cv}\brK{t_2}\,\dd t_2
\end{equation}
is the relative phase of the electron in a quantum state oscillating between the conduction and valence bands,
$\vec\xi_\mathrm{cv}(\vec k)$ is the interband matrix element, and
$E'_\mathrm{cv}\brK{t} = E'_\mathrm{c}\brK{t} - E'_\mathrm{v}\brK{t}$ is the time-dependent band gap.
Following~\cite{Kane_1960_JPCS_12_181, Argyres_1962_PR_126_1386}, we simplified our notation by introducing the modified band energies
\begin{equation}
  E'_n\brK{t} = E_n\brK{t} + e \vec F(t) \cdot \vec \xi_{nn}\brK{t}
\end{equation}
and combining the dynamic and geometric phases $\phi'_{n,\vec k} = \phi_{n,\vec k} + \gamma_{n,\vec k}$.

The probability amplitude given by Eq.~\eqref{e:ackt} is a result of the first-order adiabatic perturbation theory~\cite{Bransden_2000, Sakurai_2013}, where the role of the adiabatic parameter is played by the time-dependent crystal momentum $\vec K(t)$.
Using the Dyson series~\cite{Sakurai_2013}, it is also straightforward to derive the evolution operator and corrections of an arbitrary order (see Appendix~\ref{s:Dyson}).

Equations~\eqref{e:ackt}--\eqref{e:phicv} show that electron dynamics in strong fields emerge from a nontrivial combination of interband and intraband motion.
Their mutual influence becomes particularly clear in the Houston basis, where the probability amplitude of conduction-band occupation~\eqref{e:ackt} includes the time-dependent band gap $E'_\mathrm{cv}\brK{t}$ and interband matrix element $\vec \xi_\mathrm{cv}\brK{t}$, both depending on a crystal momentum $\vec K(t)$ that describes intraband motion.
In the subsequent text we omit, for simplicity, the geometric-phase contribution ($E'_\mathrm{cv} \rightarrow \E{cv}, \phi'_\mathrm{cv} \rightarrow \phi_\mathrm{cv}$).
We will return to its discussion in Sec.~\ref{s:Berry}.

For a monochromatic field, the integral in Eq.~\eqref{e:ackt} can be evaluated analytically using the saddle-point approximation~\cite{Keldysh_1965_JETP_20_1307} or the residue theorem~\cite{Mishima_2002_PRA_66_033401, Vanne_2007_PRA_75_033403}.
In his original paper, Keldysh used analytical expressions of energy bands and optical matrix element from the second order of the two-band $\vec k\cdot\vec p$-perturbation theory~\cite{Kane_1960_JPCS_12_181, Yu_2010}.
In this model, the band gap monotonically increases with $|\vec k|$ to infinity as
\begin{equation}\label{e:EcvKane}
  \E{cv}(\vec k) = \E{g}\left(1 + \frac{\hbar^2 k^2}{m \E{g}}\right)^{1/2},
\end{equation}
while the magnitude of the interband matrix element is estimated by
\begin{equation}\label{e:xicv}
  \xi_\mathrm{cv}\brK{t} \approx \xi_\mathrm{cv}(0) = \frac{\hbar}{2\sqrt{m \E{g}}}.
\end{equation}

The ratio of the ponderomotive and photon energies (see Table~\ref{t:gamma})
\begin{equation}
  \g{NP} = \frac{\Up}{\hbar\w0}
\end{equation}
is called the \emph{nonperturbative intensity parameter}~\cite{Reiss_1992_PQE_16_1} because it naturally appears in a perturbative expansion of strong-field theories~\cite{Keldysh_1965_JETP_20_1307, Faisal_1973_JPB_6_89, Reiss_1980_PRA_22_1786}.
At the same time, this parameter describes the number of additional photons that must be involved in the excitation of an electron-hole pair to overcome the increase of the band gap due to intraband motion [see Eq.~\eqref{e:Up}].

Cycle-averaged transitions rates derived from Eq.~\eqref{e:ackt} have particularly large values if the following energy conservation law is satisfied~\cite{Keldysh_1965_JETP_20_1307}:
\begin{equation}\label{e:EcvN}
  \overline{\E{cv}\brK{t}} = \widetilde N\hbar\w0,
\end{equation}
i.e., when exactly an integer number of photons is required to overcome the cycle-averaged band gap.
Near the $\Gamma$ point, Eq.~\eqref{e:EcvN} reduces to $\widetilde E_\mathrm{g} = \widetilde N\hbar\w0$ where, according to Eq.~\eqref{e:Up}, $\widetilde E_\mathrm{g} = \E{g} + \Up$ is the effective band gap in the presence of the driving field.
Thus the number of photons that must participate in a multiphoton transition is given by~\cite{Keldysh_1965_JETP_20_1307}
\begin{equation}\label{e:tildeN}
  \widetilde N = \left\lfloor \frac{E_\mathrm{g} + \Up}{\hbar\w0} + 1\right\rfloor,
\end{equation}
where $\lfloor x \rfloor$ denotes the floor function (the largest integer less than or equal to $x$).

\begin{figure}[t]
\includegraphics[width=0.95\linewidth]{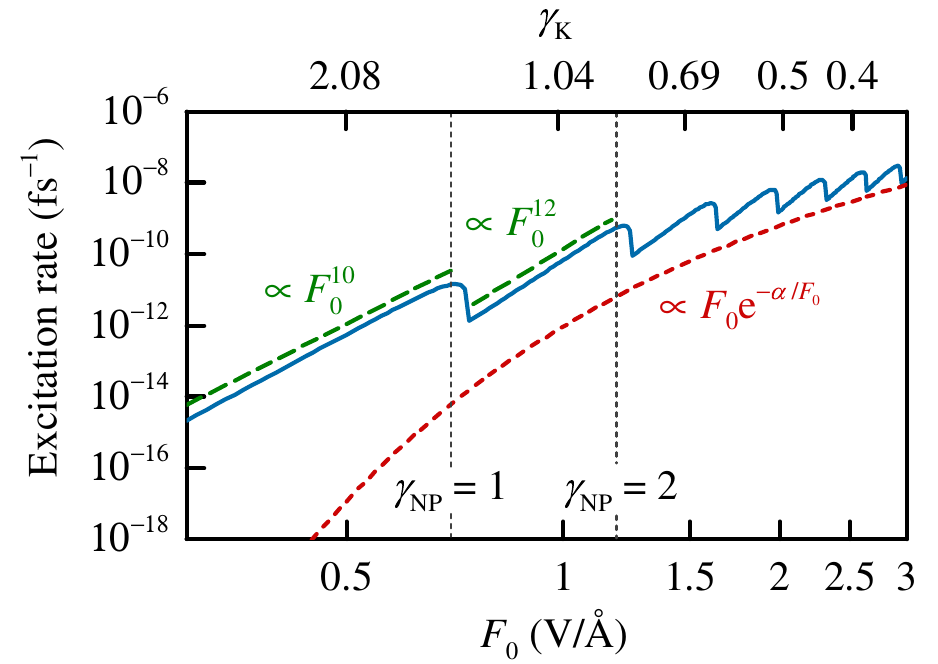}
\caption{\label{f:KeldyshSiO2}%
  (Color online) Log-log plot of the cycle-averaged excitation rate (solid blue curve) calculated using the formalism of~\textcite{Keldysh_1965_JETP_20_1307} for crystalline $\alpha$-SiO$_2$ ($\E{g} \approx 9$~eV) exposed to a monochromatic laser field with frequency $\hbar\w0 = 1.8$~eV, so that $N = 5$.
  The upper $x$-axis shows the values of the Keldysh parameter $\g{K}$ corresponding to the field amplitudes of the lower $x$-axis.
  Up to $\g{NP} \leqslant 1$, five-photon absorption is the dominant excitation mechanism, so this rate is $\propto F_0^{10}$ (dashed green line).
  The first multiphoton channel closing occurs at $\g{NP} \approx 1$, where the excitation rate first drops, and then resumes its growth a higher slope $\propto F_0^{12}$.
  For comparison, the static-field Zener tunneling rate is plotted with the short-dashed red curve.
  In the adiabatic limit, $\g{K} \ll 1$, the two rates approach each other.
}
\end{figure}

As the field amplitude $F_0$ grows, $\widetilde N$ increases stepwise.
At each of these steps, the number of photons required for multiphoton transitions increases by 1, which makes the cycle-averaged transition rate a locally decreasing function of $F_0$.
This nonperturbative phenomenon is important for $\g{NP} \gtrsim 1$ and has become known as \emph{multiphoton channel closing}~\cite{Reiss_1980_PRA_22_1786, Story_1994_PRA_49_3875, Paulus_2001_PRA_64_021401, Kopold_2002_JPB_35_217}.
We illustrate multiphoton channel closing in Fig.~\ref{f:KeldyshSiO2}, which compares the general expression for excitation rate derived by~\cite{Keldysh_1965_JETP_20_1307} with the static-field Zener tunneling rate~\cite{Zener_1934_PRSA_145_523}.
At moderately strong fields, the general excitation rate is approximately proportional to $F_0^{2 \widetilde N}$ between consecutive channel closings.
Since only an integer number of photons can be absorbed, channel closing leads to a sawtoothlike dependence of the rate on the field amplitude.
We note that, by itself, the $\g{K} \gg 1$ condition does not guarantee the applicability of the conventional perturbation theory where the small parameter is proportional to $F_0$~\cite{Reiss_1992_PQE_16_1}.
For example, Fig.~\ref{f:KeldyshSiO2} shows that the perturbative scaling law $\propto F_0^{10}$ describing multiphoton absorption starts diverging from the total excitation rate at $\g{NP} = 1$, even though, at this field strength, $\g{K} \approx 2$.

\subsection{Bloch oscillations and Wannier--Stark localization}\label{s:BO}

Once an external field drives an electron to the boundary of the Brillouin zone, the electron's de Broglie wavelength becomes equal to twice the lattice period:
\begin{equation*}
  |\vec K(\vec k = 0, t)| = \frac{\pi}{a} = \frac{2\pi}{\lambda} \Rightarrow \lambda = 2a,
\end{equation*}
which is a condition for Bragg scattering on the lattice potential.
In the reduced zone scheme, an electron's trajectory terminates at the boundary of the first Brillouin zone and continues on the opposite side of the zone.
If $\vec K(t)$ traverses the Brillouin zone several times per optical cycle, then an electron confined to a particular band is said to perform Bloch oscillations.
In real space, Bragg scattering of an electron wavepacket rapidly changes its group velocity according to Eq.~\eqref{e:Bloch_r}.
Multiple coherent scattering events of this type reduce the wavepacket's width and displacement, which is called Wannier--Stark localization.

The probability of Bragg scattering is small if the driving field is polarized along a crystallographic direction where the lattice potential is particularly weak.
In this case, the band gap between the first and second conduction (valence) bands is small\footnote{If the gap is zero, it is convenient to combine the degenerate bands into a single band in the extended BZ.}, and the tunneling probability between them is close to 1.
However, for a particular band, there are always crystallographic directions where this degeneracy is lifted by a lattice potential, so that the band becomes isolated from the others and has a finite bandwidth $\Delta_n$.

\begin{figure}[t]
\includegraphics[width=\linewidth]{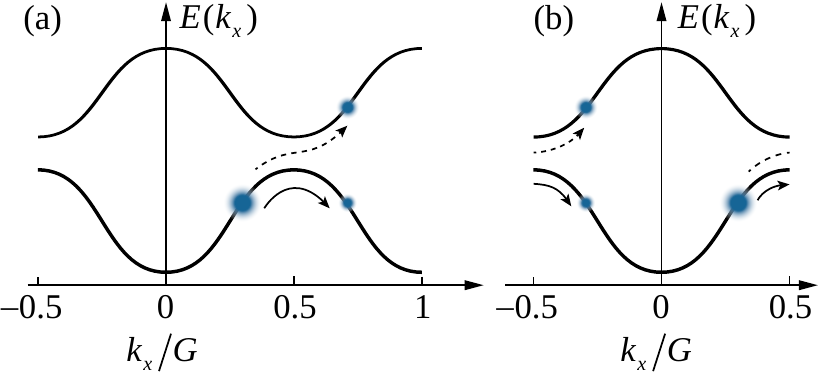}
\caption{\label{f:Bands}%
  (Color online) Interplay between Bloch oscillations and interband tunneling of charge carriers in the extended~(a) and reduced~(b) zone schemes.
  Solid arrows depict the intraband motion of carriers leading to Bloch oscillations, and dashed arrows depict interband tunneling.
  Here, $G = 2\pi/a$ denotes the BZ width, and $a$ is the lattice constant.
}
\end{figure}

If an electron's reciprocal-space trajectory passes near an avoided crossing between two energy bands, then there is a nonzero interband transition probability.
The stronger is the electric field, the faster $\vec K(t)$ changes, increasing the probabilities of such transitions.
Figure~\ref{f:Bands} schematically illustrates the transition of an electron wavepacket through a BZ boundary, where two bands have a small gap between them.
The Landau--Zener dynamics are best illustrated in the extended zone scheme, which we use for Fig.~\ref{f:Bands}(a).
The same dynamics in the reduced zone scheme are illustrated in Fig.~\ref{f:Bands}(b).

The Keldysh theory was initially developed for bands approximated by Eq.~\eqref{e:EcvKane}, which implies that the probability of coherent scattering on the lattice potential is negligibly small.
However, it is not negligible in real solids, and even a small fraction of Bragg-reflected electrons can lead to detectable outcomes, where high-harmonic generation is a prominent example~\cite{Ghimire_2011_NP_7_138}.
Analytical approaches to strong-field dynamics beyond the effective mass approximation were developed by many, including~\cite{Krieger_1986_PRB_33_5494, Rotvig_1995_PRL_74_1831, Gruzdev_2007_PRB_75_205106, Hawkins_2013_PRA_87_063842, Zhokhov_2014_PRL_113_133903, McDonald_2017_PRL_118_173601}.
Also, a purely adiabatic evolution was considered in, e.g.,~\cite{Wannier_1962_RMP_34_645, Dunlap_1986_PRB_34_3625, Holthaus_1992_PRL_69_351}.

The physical interpretation of several phenomena that we discuss becomes particularly clear in real space.
For a linearly polarized field, it is advantageous to limit the real-space analysis to the direction along the field, so that we do not lose advantages of the reciprocal-space representation for directions orthogonal to the field.
This is possible in the \emph{hybrid representation}~\cite{Argyres_1962_PR_126_1386, Sgiarovello_2001_PRB_64_115202, Marzari_2012_RMP_84_1419} where the wavefunctions and operators are transformed to the coordinate space only over the component of crystal momentum pointing along the field polarization.
Let this direction be the $x$ axis.
In the hybrid representation~\cite{Fritsche_1966_PSSB_13_487}, a Houston function~\eqref{e:Houston} can be expanded into a series of functions known as Kane states~\cite{Kane_1960_JPCS_12_181, Fritsche_1966_PSSB_13_487, Glutsch_2004}:
\begin{equation}\label{e:HoustonKane}
  \psi_{n, \vec k}^\mathrm{(H)}(\vec r, t) = \frac{1}{\sqrt G} \sum_{l = -\infty}^{+\infty} \varphi_{n, l}^\mathrm{(K)}(\vec k_\perp, \vec r) \e^{-\frac{\ii}{\hbar} \mathcal E_{n, l}^\mathrm{(K)}(\vec k_\perp) t},
\end{equation}
where $l$ enumerates lattice sites, $\vec k_\perp$ is the part of the crystal momentum perpendicular to the field, and $G = 2\pi/a$ denotes the BZ width in the $k_x$ direction.
Equation~\eqref{e:HoustonKane} shows that the Houston state $\psi_{n, \vec k}^\mathrm{(H)}$ is a generating function of the Kane states $\varphi_{n,l}^\mathrm{(K)}$.
Alternatively, the hybrid Kane functions can be derived from the solutions of an adiabatic eigenproblem in the length gauge and in the single-band approximation.
This is rigorously discussed in Appendix~\ref{s:Kane}.

\begin{figure}[t]
  \includegraphics[width=\linewidth]{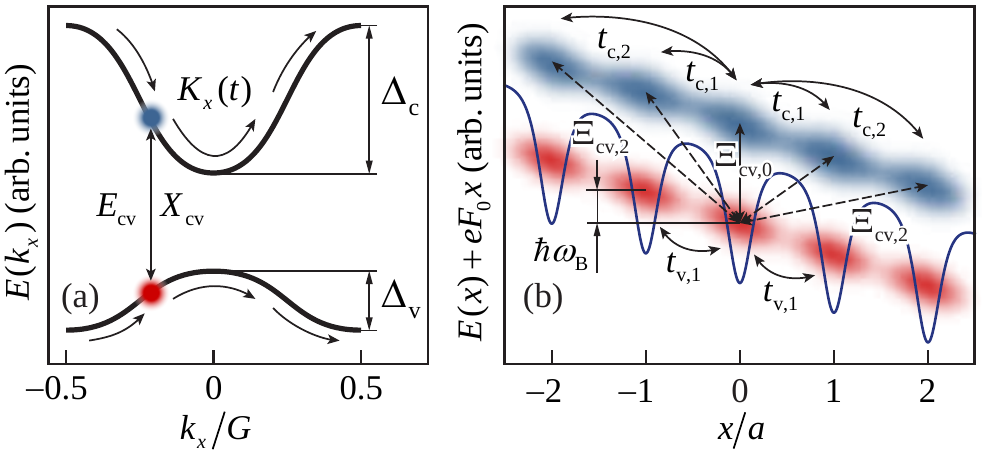}
  \caption{\label{f:HoustonKane}%
    (Color online) Two equivalent representations of strong-field electron dynamics in periodic potentials.
    (a)~Reciprocal-space picture in the Houston basis.
    The motion of an electron-hole pair is determined by inter- and intraband transitions, which depend on the time-dependent energy gap $E_\mathrm{cv}\brKx{t}$, optical matrix element $X_\mathrm{cv}(K_x(t)) \equiv \vec e_x\cdot \vec\xi_\mathrm{cv}\brK{t}$, and crystal momentum $K_x(t)$.
    (b)~Real-space picture in the Kane basis.
    Electronic states are separated by the Bloch energy $\hbar\w{B} = |e F_0| a$.
    Solid arrows depict the intraband motion due to tunneling between lattice sites, while the dashed arrows represent the interband transitions.
    Their probabilities are determined by the hopping integrals $t_{n,\ell}$ and the real-space optical matrix elements $\Xi_{\mathrm{cv},\ell}$, respectively.
    Widths of the energy bands are given by the sum over hopping integrals: $\Delta_n = \abs{4\sum_{\ell = 1}^{\infty} t_{n, \ell}}$.
  }
\end{figure}

The energies of Kane states are given by
\begin{equation}\label{e:KaneE}
  \mathcal E_{n,l}^\mathrm{(K)}(\vec k_\perp) = \overline{E}'_n(\vec k_\perp) + \frac{2\pi l |e F_0|}{G}.
\end{equation}
The right-hand side of Eq.~\eqref{e:KaneE} consists of the continuous part $\overline{E}'_n(\vec k_\perp)$ [see Eq.~\eqref{e:oEn}] and the discrete components $2\pi l eF_0/G = l e a F_0 = \hbar l\w{B}$, separated by multiples of Bloch frequency and known as the Wannier--Stark ladder.

Representations of electron dynamics in the Houston and Kane bases are schematically shown in Fig.~\ref{f:HoustonKane}.
Each Kane state is a time-independent function; dynamics emerge once we consider their superpositions.
While the coherent superposition of an infinite number of Kane states forms a delocalized Houston function [see Eq.~\eqref{e:HoustonKane}], a finite number of Kane states forms a spatially localized electron wavepacket.
In both cases, the result is a temporally periodic wavefunction that oscillates with the Bloch frequency.
Figure~\ref{f:HoustonKane}(a) illustrates these oscillations in reciprocal space.
In real space, a localized wavepacket oscillates within the range occupied by the selected Kane states [see Fig.~\ref{f:HoustonKane}(b)].
Because of the spatial overlap of these states, the width of this wavepacket can be smaller than the range of its motion.

\begin{figure}[t]
  \includegraphics[width=\linewidth]{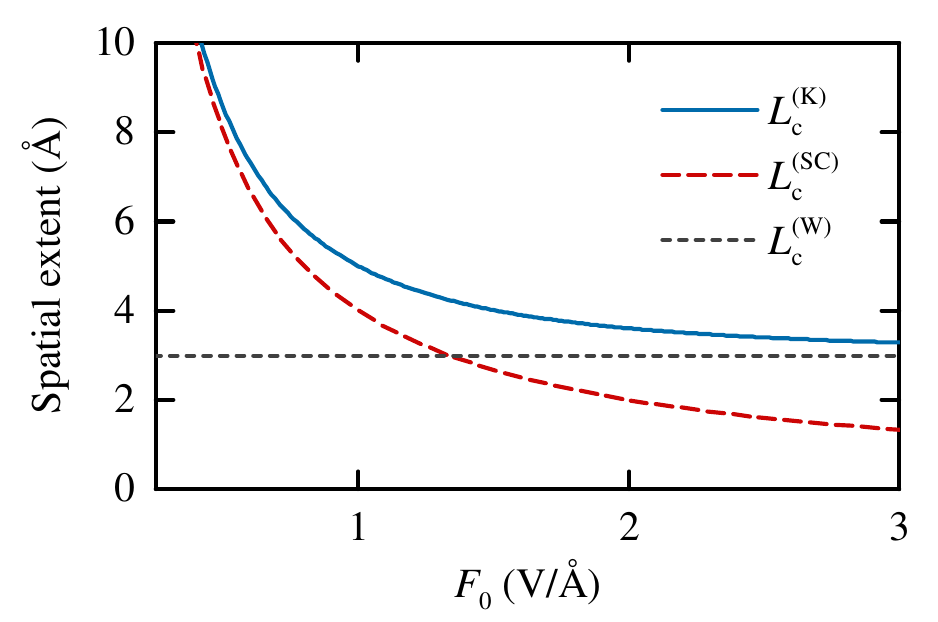}
  \caption{\label{f:Lc}%
    (Color online) Spatial extent of an electron in the first conduction band of $\alpha$-SiO$_2$ as a function of the field calculated according to fully quantum-mechanical~\eqref{e:LnK} (solid blue curve) and semiclassical models~\eqref{e:LnSC} (dashed red curve).
    The spatial extent of the Kane function $L\mathrm{_c^{(K)}}$ approaches the extent of a maximally localized Wannier function $L\mathrm{_c^{(W)}}$ (short-dashed black line) in the strong-field limit.
    Here we assumed the following material parameters
    $\Delta_\mathrm{c} \approx 3.3$~eV~\cite{Kresse_2012_PRB_85_045205}, and $L_\mathrm{c}^\mathrm{(W)} \approx 3$~\AA{}~\cite{Mustafa_2015_PRB_92_165134}.
  }
\end{figure}

The spatial extent of a Kane state in the $n$th band is given by~\cite{Glutsch_1999_JP_11_5533, Glutsch_2004}
\begin{equation}\label{e:LnK}
  L_n^\mathrm{(K)} = \sqrt{\bigl[L_n^\mathrm{(SC)}\bigr]^2 + \bigl[L_n^\mathrm{(W)}\bigr]^2}.
\end{equation}
Figure~\ref{f:Lc} shows that in the weak-field limit Eq.~\eqref{e:LnK} approaches
\begin{equation}\label{e:LnSC}
  L_n^\mathrm{(SC)} = \frac{\Delta_n}{|e F_0|},
\end{equation}
which is a well-known semiclassical (SC) formula used by many authors~\cite{Mendez_1988_PRL_60_2426, Voisin_1988_PRL_61_1639, Dignam_1994_PRB_49_10502, Schiffrin_2013_Nature_493_70}.
In the strong-field limit $F_0 \rightarrow\pm\infty$, Eq.~\eqref{e:LnK} asymptotically approaches the extent of a maximally localized Wannier function $L_n^\mathrm{(W)}$~\cite{Glutsch_1999_JP_11_5533, Marzari_2012_RMP_84_1419}.

Interband transitions in static or slowly varying fields can be interpreted, in real space, as transitions between Kane states.
Indeed, each Kane function is formed from the Bloch states of a particular band [see Eq.~\eqref{e:Kane}], so interband dynamics can be described as transitions between subsets of Kane functions $\varphi_{n, l}^\mathrm{(K)} \equiv |n, l\rangle$ corresponding to different bands.
The probabilities of these transitions are determined by the real-space optical matrix elements $\Xi_{nm,\ell}(F_0) = \langle n, \ell| x | m, 0\rangle$ depending on the electric field.
Stronger fields allow interband tunneling between closer sites, but they also make Kane functions more localized, which tends to reduce $|\Xi_{nm, \ell}(F_0)|$.

Once a static electric field reaches a strength for which a pair of valence- and conduction-band Kane states localized at different lattice sites has the same energy, adiabatic tunneling between them becomes allowed by the energy conservation law.
This occurs for $\ell\hbar\w{B}/\E{g} = \ell\g{BH} = 1$, that is, $\ell a |e F_0| = \E{g}$, where $\ell = |l_1 - l_2| = 1, 2, ...$ denotes the distance between lattice sites.
We refer to $\g{BH}$, introduced in Table~\ref{t:gamma}, as the \emph{band hybridization} parameter.
Note that, in time-dependent fields, purely adiabatic transitions coexist with diabatic tunneling and multiphoton processes; therefore realistic simulations of strong-field phenomena for ultrashort laser pulses must include all possible excitation mechanisms.

\begin{figure}[t]
  \includegraphics[width=\linewidth]{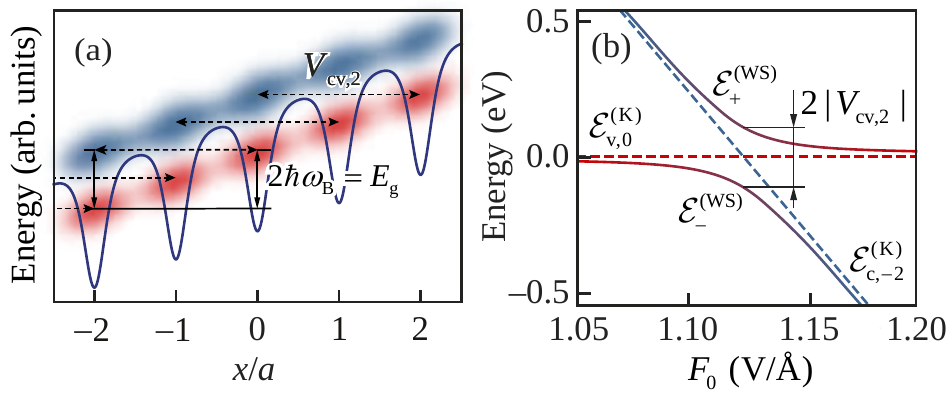}
  \caption{\label{f:WS-Kane}%
    (Color online) (a)~Lattice potential in an external field at $2\g{BH} = 1$, where adiabatic tunneling to the second neighbor is allowed.
    (b)~Wannier--Stark ladders in one- and two-band approximations as a function of the electric-field amplitude.
    Dashed lines depict the energies of one-band (diabatic) Kane states $\mathcal E_{n, \ell}^\mathrm{(K)}$, and solid curves show the energies of two-band (adiabatic) Wannier--Stark states $\mathcal E_{\pm}^\mathrm{(WS)}$.
  }
\end{figure}

Wannier--Stark ladders for coupled bands can be approximated by the roots of the following secular equation~\cite{Bastard_1994_PRB_50_4445, Grecchi_1995_AP_241_258}:
\begin{equation}\label{e:VnmE}
  \det(V_{nm,\ell} - \mathcal E \delta_{nm}) = 0,
\end{equation}
where $V_{nm,\ell} = e F_0 \Xi_{nm,\ell}$ is the interaction term describing interband transitions.
In the simplest case of two bands, the analytical solution is well known:
\begin{multline}\label{e:EWS}
\mathcal E_{\pm,\ell}^\mathrm{(WS)} = \frac{1}{2}\left(
\mathcal E_{\mathrm{c},i}^\mathrm{(K)}
+ \mathcal E_{\mathrm{v},i+\ell}^\mathrm{(K)}
\right)
\\
\pm\frac{1}{2}\sqrt{\left(
  \mathcal E_{\mathrm{c},i}^\mathrm{(K)}
  - \mathcal E_{\mathrm{v},i+\ell}^\mathrm{(K)}
  \right)^2 + 4 |V_\mathrm{cv, \ell}|^2},
\end{multline}
where the energy gap at an anticrossing is given by $2 |V_\mathrm{cv, \ell}|$ [see Fig.~\ref{f:WS-Kane}(b)].

\begin{figure}[t]
  \includegraphics[width=\linewidth]{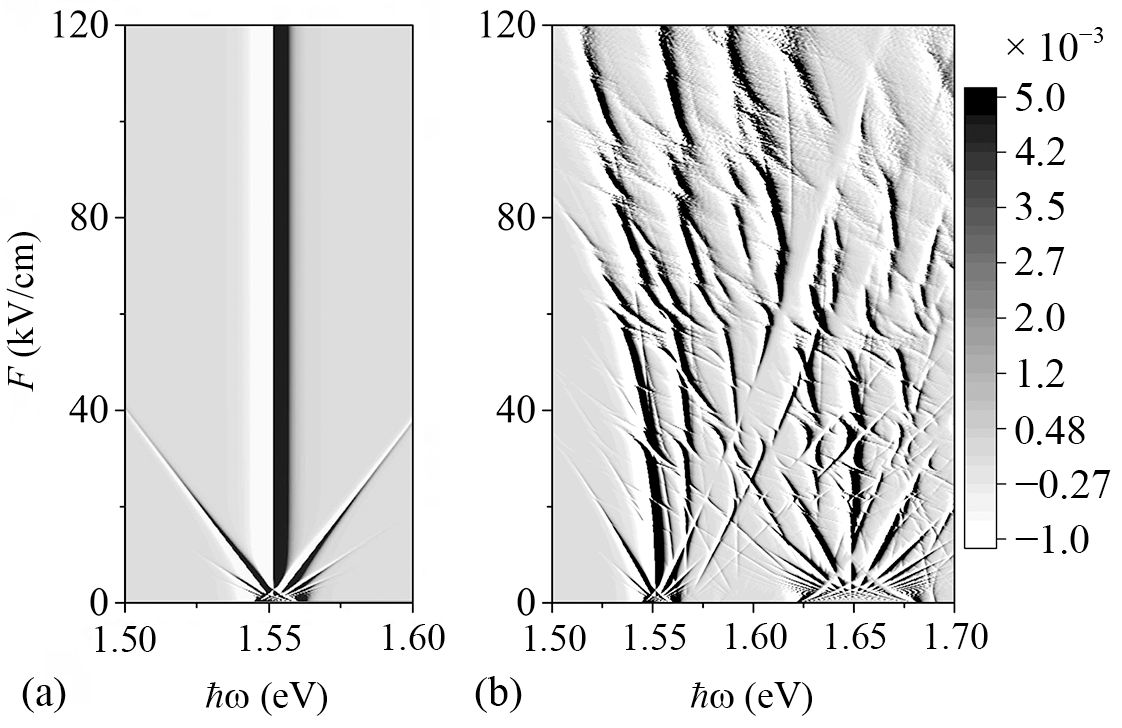}
  \caption{\label{f:Kane-WS}%
    Derivative of the optical density of states $\dd D/\dd\omega$ in the GaAs/GaAlAs superlattice as a function of the transition energy $\hbar\omega$ and the field strength $F$.
    (a) The one-band (Kane) approximation.
    (b) The Wannier--Stark ladder calculated numerically including a few minibands and transitions between them.
    From~\cite{Glutsch_1999_PRB_60_16584}.
  }
\end{figure}

Accounting for transitions between bands when solving the adiabatic eigenproblem removes degeneracies in the Wannier--Stark ladder, turning them into anticrossings [Fig.~\ref{f:WS-Kane}(b) and Fig.~\ref{f:Kane-WS}(b)]~\cite{Avron_1982_AP_143_33, Leo_2003, Glutsch_2004}.
Under the assumption that an electron is confined to a certain finite subset of energy bands, the adiabatic wave functions obtained by solving the corresponding eigenproblem are known as Wannier--Stark states.
Within this basis, the adiabatic transition through an avoided crossing transforms a state that is mainly constructed from the wave functions of one band into a state that is predominantly formed by the wavefunctions of another band.
Therefore, the adiabatic transitions [along the solid lines in Fig.~\ref{f:WS-Kane}(b)] in the two-band Wannier--Stark basis represent interband tunneling~\cite{Apalkov_2012_PRB_86_165118, Apalkov_2015}, while diabatic transitions (along the dashed lines) correspond to intraband motion.
Note that this is opposite to the picture of carrier dynamics in the single-band bases of Houston or Kane functions, where adiabatic evolution means staying in the same energy band (see Fig.~\ref{f:HoustonKane}).

Wannier--Stark states are always defined for a limited subset of bands $\mathbf{n} = \{n_1, n_2, \ldots, n_{\max} \}$.
Any transitions outside this subset turn these states into metastable resonances with a complex-valued energy spectrum $\mathcal E_{\mathbf{n}, \ell}^\mathrm{(WS)} - \ii \Gamma_{\mathbf{n}, \ell}/2$, where $\Gamma_{\mathbf{n}, \ell}$ is the rate at which the population is transferred to other bands.
In spectroscopic measurements, the real and imaginary parts of Wannier--Stark resonances correspond to the spectral positions and linewidths, respectively~\cite{Avron_1976_PRL_37_1568, Nenciu_1991_RMP_63_91, Gluck_1999_PRL_83_891, Rosam_2003_PRB_68_125301}.
This is demonstrated in Fig.~\ref{f:Kane-WS}, where the single- and multiple-band calculations for semiconductor superlattices are compared.
In stronger fields, the linewidths increase, which is a signature of eigenfunction delocalization and transitions to higher bands due to the Zener breakdown~\cite{Glutsch_1999_PRB_60_16584}.
In general, Wannier--Stark resonances can be calculated from the complex poles of a rigorously constructed $S$-matrix~\cite{Gluck_2002_PR_366_103} or via diagonalization of a non-Hermitian Hamiltonian~\cite{Moiseyev_2011}.
To the best of our knowledge, ab initio calculations of Wannier--Stark resonances have so far been reported only for artificial periodic structures and model potentials, rather than real solids.

In a periodic potential exposed to a static electric field, Wannier--Stark localization naturally occurs if an electron remains in a subset of bands~\cite{Wannier_1962_RMP_34_645}.
For an oscillating linearly polarized field, the spatial extent of a localized wavepacket usually increases over each period.
Indeed, Wannier--Stark localization might occur at the crests of the field, but it plays no role at its zero crossings.
Nevertheless, nonspreading wave packets may exist even in oscillating fields: the displacement and spatial extent of such a wavepacket oscillate within certain limits.
This phenomenon is known as \emph{dynamic localization}.
It was predicted by~\textcite{Dunlap_1986_PRB_34_3625}, who used a single-band nearest-neighbor tight-binding model and found solutions of the TDSE where wavepackets remain localized in a direction along the field polarization.
For sinusoidal fields, these solutions exist if the ratio $\g{DL} = \w{B}/\w0$ is one of the roots of the zeroth order Bessel function $J_0$ (the first zero of the function is at $\approx 2.405$); consequently, no localization occurs if $\g{DL} \ll 1$.
Hence, $\g{DL}$ is known as the dynamic localization parameter (see Table~\ref{t:gamma}).

In real crystals, hopping between distant neighbors and interband transitions prevent perfect localization.
Nevertheless, the nearest-neighbor hopping is dominant in most of systems, so a significant reduction of a wavepacket spread takes place when $\g{DL}$ is a root of $J_0$, a less significant reduction takes place when $2\g{DL}$ is a root of $J_0$, and so on~\cite{Dunlap_1986_PRB_34_3625}.

The parameter $\g{DL}$ also describes the nonlinearity of a wave packet's group velocity, and thus it plays an essential role in the high-field transport phenomena and high-harmonic generation in superlattices and solids~\cite{Tsu_1971_APL_19_246, Ignatov_1976_PSSB_73_327, Pronin_1994_PRB_50_3473, Feise_1999_APL_75_3536, Wegener_2005, Golde_2008_PRB_77_075330, Ghimire_2011_NP_7_138, Luu_2015_Nature_521_498, Hammond_2017_NP_11_594}.
In particular, it appears in the cutoff condition
\begin{equation}\label{e:Ncutoff}
  N_\mathrm{cutoff} = \ellmax \g{DL}
\end{equation}
for the high-frequency plateaus generated by the intraband motion of electrons.
In the tight-binding approximation, $\ellmax$ is the maximal number of distant neighbors (spatial harmonics) contributing to the band dispersion.
According to recent measurements on wide-band-gap materials (ZnO, SiO$_2$) exposed to IR pulses, $N_\mathrm{cutoff}$ may extend up to 25 orders involving about $\ellmax = 6$ neighbors~\cite{Ghimire_2011_NP_7_138, Luu_2015_Nature_521_498, Garg_2016_Nature_538_7625}.

Let us illustrate two opposite limits defined by $\g{DL}$ with a simple one-dimensional single-band model~\cite{Pronin_1994_PRB_50_3473, Feise_1999_APL_75_3536, Golde_2008_PRB_77_075330, Muecke_2011_PRB_84_081202, Ghimire_2011_NP_7_138, Luu_2015_Nature_521_498}.
An electron wavepacket with a large spatial extent has a small spread of crystal momenta.
Thus it is easier to evaluate its group velocity and displacement using the acceleration theorem~\eqref{e:Bloch_k}, without decomposing the wavepacket into a large number of Kane states.

The eigenvalues of the tight-binding Hamiltonian with distant-neighbor hopping~\cite{Stockhofe_2015_PRA_91_023606},
\begin{multline}\label{e:HTB}
  \hat H_\mathrm{TB} = \sum_{n}\sum_{i =-\infty}^\infty \biggl[ V_i \hat a_{n,i}^\dagger \hat a_{n,i}
\\
- \sum_{\ell = 1}^{\ell_\mathrm{max}} t_{n, \ell}
  \left( \hat a_{n,i}^\dagger \hat a_{n,i+\ell} + \hat a_{n,i+\ell}^\dagger a_{n,i}\right)
  \biggr]
\end{multline}
form the energy bands
\begin{equation}\label{e:En}
  E_n(k_x) = \sum_{\ell=0}^{\ell_\mathrm{max}} \varepsilon_{n,\ell} \cos(k_x \ell a),
\end{equation}
where $\varepsilon_{n, \ell} \equiv -2 t_{n, \ell}$ are hopping integrals between the Wannier states separated by $\ell$ lattice constants,
$\varepsilon_{n, 0}$ is a band offset,
$\hat a_i^\dagger$ and $\hat a_i$ denote creation and annihilation operators, and
$V_i$ is a local scalar potential at site $i$.

The instantaneous group velocity of an electron wavepacket driven by a strong field in the band with dispersion~\eqref{e:En} is given by~\cite{Luu_2015_Nature_521_498}
\begin{equation}\label{e:vnKx}
  v_n\brKx{t} = \frac{1}{\hbar} \left.\frac{\partial E_n}{\partial k_x}\right|_{K_x(t)}= -\frac{1}{\hbar} \sum_{\ell=0}^{\ell_\mathrm{max}} \ell \varepsilon_{n, \ell} \sin\bm{(}K_x(t) \ell a\bm{)}.
\end{equation}

By integrating the group velocity over time, one obtains the relative displacement of the wave packet's center of mass:
\begin{equation}\label{e:DxnKx}
  \Delta x_n\bm{(}K_x(t)\bm{)} = \int_{t_0}^t v_n\brKx{t_1}\,\dd t_1.
\end{equation}

\begin{figure}[!ht]
  \includegraphics[width=\linewidth]{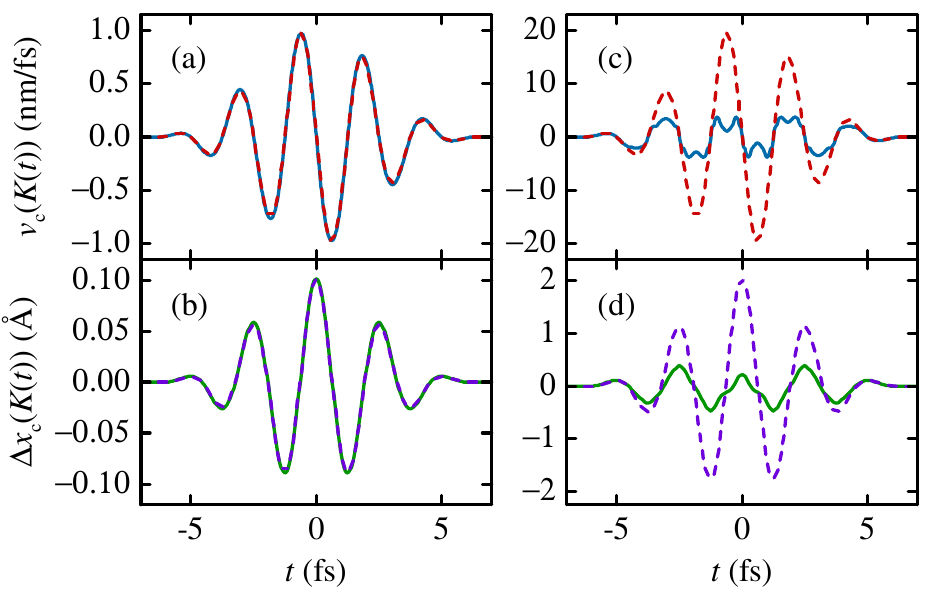}
  \caption{\label{f:vc-xc}%
    (Color online) Instantaneous group velocity $v_\mathrm{c}$ and relative displacement $\Delta x_\mathrm{c}$ of an electron wavepacket driven by (a), (b) weak and (c), (d) strong laser pulses in the first conduction band of SiO$_2$ from $\vec k = 0$ along the $\Gamma$--M direction.
    Dashed curves depict the results obtained within the effective-mass approximation [Eqs.~\eqref{e:vnEMA} and~\eqref{e:DxnEMA}], and solid curves correspond to the entire band dispersion [Eqs.~\eqref{e:vnKx} and~\eqref{e:DxnKx}].
    In the weak-field limit ($F_0 = 0.05$~V/\AA{}, $\g{DL} \approx 0.15$), both expressions coincide, but in the strong field ($F_0 = 1$~V/\AA{}, $\g{DL} \approx 3$) Eqs.~\eqref{e:vnKx} and~\eqref{e:DxnKx} significantly deviate from the EMA result and demonstrate nonlinearities due to dynamic Bloch oscillations.}
\end{figure}

If $\ell \g{DL} \ll 1$, the argument of the sine function in Eq.~\eqref{e:vnKx} is small, and the $\sin x \approx x$ approximation yields the following expression for the group velocity of carriers in the $n$th band:
\begin{equation}\label{e:vnEMA}
  v_n^\mathrm{(EMA)}\brKx{t} = \frac{\hbar}{m_n} K_x(t)
= \frac{1}{m_n}[\hbar k_x + e A(t)].
\end{equation}
The increment of the velocity is proportional to the vector potential \eqref{e:A} [see Figs.~\ref{f:vc-xc}(a), (b)].
The corresponding relative displacement is given by
\begin{equation}\label{e:DxnEMA}
  \Delta x_n^\mathrm{(EMA)}\brKx{t}
  = \frac{1}{m_n}\left[
    \hbar k_x (t - t_0) + e \int_{t_0}^{t} A(t_1)\,\dd t_1
  \right].
\end{equation}
These expressions can also be obtained in the EMA, where the band dispersion law is given by Eq.~\eqref{e:EnkEMA}.

In the opposite case of $\ell \g{DL} \gg 1$, one recovers the quasistatic-field limit for the $\ell$th distant neighbor, where a field-driven electron confined to a particular band performs a few Bloch oscillations per optical cycle.
Hereafter, we will refer to Bloch oscillations driven by a time-dependent field in this regime as \emph{dynamic Bloch oscillations}.
Figures~\ref{f:vc-xc}(c), (d) show that wavepacket's group velocity and relative displacement are much smaller than those predicted by the EMA.
The decelerated intraband motion results in a negative differential conductance~\cite{Esaki_1970_IJRD_14_61, Tsu_1971_APL_19_246, Tsu_2011} and the emission of high-energy photons with frequencies up to $\ell_\mathrm{max}\w{B}$ in HHG experiments~\cite{Golde_2008_PRB_77_075330, Ghimire_2011_NP_7_138, Luu_2015_Nature_521_498}.

The ratio of Bloch and ponderomotive energies gives the parameter $\g{BP}$ (see Table~\ref{t:gamma}), which we include into our classification scheme for completeness, but for which we do not currently have a clear physical example illustrating its significance.
In general terms, this parameter describes the balance between the field-induced intraband motion, which tends to delocalize an electron wavepacket, and Bloch oscillations, which tend to localize it.

\subsection{Interaction with a nearly-resonant field}
\label{s:Rabi}

A sufficiently strong laser field with a carrier frequency close to resonance with the band gap of a solid may transfer population back and forth between the valence and conduction bands.
These Rabi oscillations rely on coherence between the involved quantum states and hence can occur only on time scales shorter than the decay of quantum coherence.
The peak frequency of these oscillations is given by Eq.~\eqref{e:wR}.

In the resonant case, the adiabaticity parameter distinguishing interaction regimes is the ratio of the instantaneous Rabi frequency $\w{R}$ at the peak of the laser pulse to the laser frequency $\w0$ (see Table~\ref{t:gamma}):
\begin{equation*}
  \g{RF}^{(0)} = \frac{\w{R}}{\w0}.
\end{equation*}
This parameter counts the number of Rabi cycles within one laser field oscillation.
The weak-field limit $\g{RF}^{(0)} \ll 1$ corresponds to the envelope Rabi flopping (ERF); $\g{RF}^{(0)} \gg 1$ is the tunneling limit~\cite{Takagahara_2003, Kaertner_2004}; at $\g{RF}^{(0)} \sim 1$, these two pictures merge in the intermediate regime known as carrier-wave Rabi flopping (CWRF)~\cite{Hughes_1998_PRL_81_3363}.
Alternatively, one can distinguish these limits by taking the ratio of the Rabi and transition frequencies $\g{RF}^\mathrm{(g)} = \hbar\w{R}/\E{g}$.
Ratios $\g{RF}^\mathrm{(0)}$ and $\g{RF}^\mathrm{(g)}$ naturally appear as small parameters of the conventional perturbation theory constructed in the field-free basis~\cite{Lamb_1987_PRA_36_2763, Aversa_1995_PRB_52_14636, Boyd_2013}.

We start our discussion with a brief review of the two-level model.
These results can be directly generalized for a two-band model of periodic systems in a single-particle approximation, but only for moderate intensities at which the intraband motion is still negligible.
In the weak-field limit, an analytical solution of the two-level TDSE or Bloch equations for the density matrix can be obtained in a closed analytical form within the rotating-wave approximation (RWA).
Initially, this approach was employed in the Rabi theory of a nuclear magnetic resonance~\cite{Rabi_1936_PR_49_324} and then in the Jaynes--Cummings model~\cite{Janes_1963_PIEEE_51_89} describing a two-level atom interacting with a single mode of an optical cavity.

For a monochromatic field
\begin{equation*}
  \vec F(t) = \frac{\vec F_0}{2} \left(\e^{\ii\w0 t} + \e^{-\ii\w0 t} \right),
\end{equation*}
the probability amplitudes in the two-level model form the following system of two equations:
\begin{align*}
  \ii\hbar\frac{\dd a_1}{\dd t} &= \frac{e}{2} \vec F_0 \cdot \vec r_{12} \left[
  \e^{-\ii(\w0 + \w{21})t} + \e^{\ii(\w0 - \w{21})t}
  \right] a_2
,\\
  \ii\hbar\frac{\dd a_2}{\dd t} &= \frac{e}{2} \vec F_0 \cdot \vec r_{21} \left[
    \e^{-\ii(\w0 - \w{21})t} + \e^{\ii(\w0 + \w{21})t}
  \right] a_1,
\end{align*}
where $\w{21} = -\w{12}$ is the transition frequency, and $\vec r_{12}$ is the optical matrix element.
Near the resonance, where $\Delta = \w{21} - \w{0} \ll \w{0}$, the corotating (resonant) terms $\sim \e^{\pm\ii(\w0 - \w{21})t}$ oscillate much slower than the laser field, while the counterrotating (nonresonant) ones $\sim \e^{\pm\ii(\w0 + \w{21})t}$ oscillate much faster and can be neglected at weak intensities~\cite{HaugKoch_2009, Allen_2012}.
Introducing the population inversion $w(t) = |a_2(t)|^2 - |a_1(t)|^2$, one obtains the classical result of Rabi, $w(t) = -\cos(\tW{R} t)$, showing that the population oscillates between the two states with the generalized envelope Rabi frequency $\tW{R} = \sqrt{\tw{R}^2 + \Delta^2}$, where $\tw{R} = |e \vec F_0 \cdot \vec r_{12}|/\hbar$.

RWA can also be considered as a special case of the adiabatic eigenproblem.
Diagonalization of the RWA Hamiltonian yields adiabatic eigenstates that are shifted and split in comparison to the field-free atomic levels~\cite{HaugKoch_2009}
\begin{equation}\label{e:Epm}
  \mathcal E_{1,\pm} = E_1 + \frac{\hbar}{2}(\Delta \pm \tW{R})
,\quad
  \mathcal E_{2,\pm} = E_2 - \frac{\hbar}{2}(\Delta \mp \tW{R})
.
\end{equation}

\begin{figure}[!ht]
  \includegraphics[width=\linewidth]{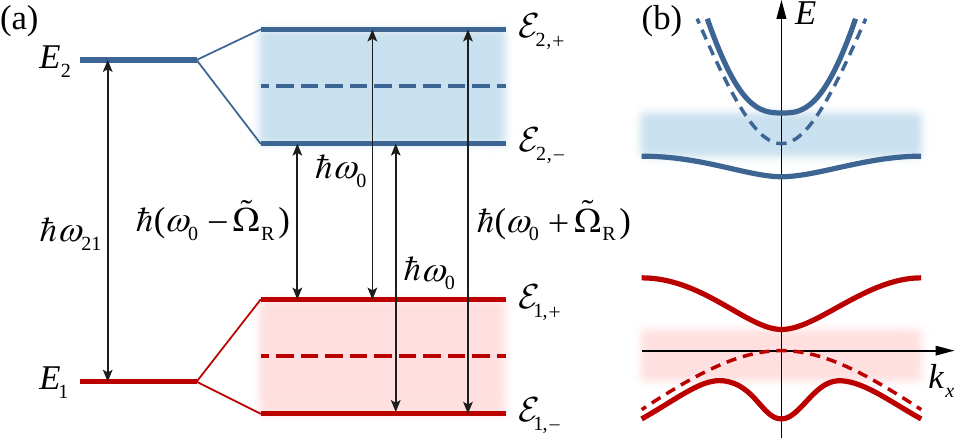}
  \caption{\label{f:DynStark}%
    (Color online) Schematic drawing of level splitting in the (a) two-level system and in a (b) solid excited by a nearly-resonant strong field and described within the RWA.
    (a) Possible optical transitions.
    Long dashed lines denote the Stark shifts of unperturbed states by $\pm \hbar\Delta/2$.
    Short dashed lines denote the field-free bands, and the shaded areas show the light-induced gaps.
  }
\end{figure}

Renormalized energy levels described by Eq.~\eqref{e:Epm} can be observed as modifications of an absorption or emission spectrum known as the \emph{Autler--Townes} or \emph{dynamic Stark effect}~\cite{Autler_1955_PR_100_703, Delone_1999_PU_42_669}.
The resulting spectrum of an atom shows the \emph{Mollow triplet}~\cite{Wu_1977_PRL_38_1077} consisting of the peak at $\w0$ and two sidebands at $\w0 \pm \tW{R}$ (see Fig.~\ref{f:DynStark}).
It can also be regarded as the resonant analog of self-phase modulation~\cite{Wegener_2005}.
This triplet was experimentally observed for discrete levels of an atom, as well as for energy bands of a semiconductor resonantly excited by an intense ultrashort laser pulse~\cite{Vu_2004_PRL_92_217403}.

\begin{figure}[t]
  \includegraphics[width=\linewidth]{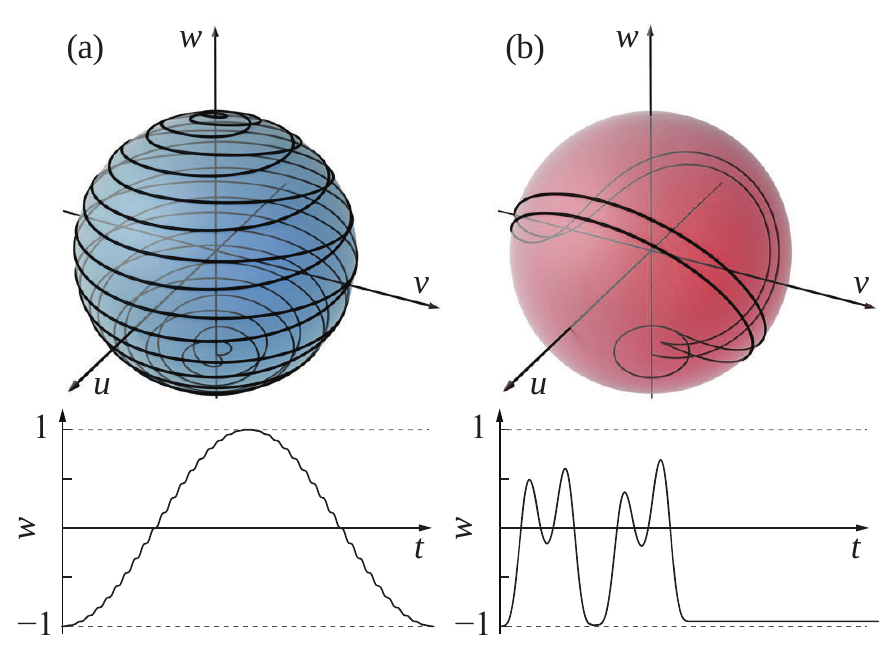}
  \caption{\label{f:RabiFlopping}%
    (Color online) .
    Evolution of (upper plots) Bloch vector and (lower plots) population inversion in the envelope- and carrier-wave Rabi flopping regimes of a two-level system~\cite{Muecke_2001_PRL_87_057401}.
    Here, $u(t) = 2\Re(a_1^*a_2)$, $v(t) = 2\Im(a_2^*a_1)$, and $w(t) = |a_2|^2 - |a_1|^2$ are the components of the Bloch vector.
    (a)~Envelope Rabi flopping (ERF) in the weak field ($\g{RF} = 0.05$) for a $2\pi$-pulse.
    Numerically calculated inversion $w(t)$ closely reproduces the RWA result $w(t) = -\cos(\tW{R} t)$.
    It is modulated by the Bloch--Siegert oscillations due to counter-rotating term omitted in the RWA.
    (b)~In the CWRF regime ($\g{RF} \sim 1$), the contribution from the counter-rotating term becomes significant, and the two-cycle pulse causes oscillations of inversion with a period comparable to that of an optical oscillation $T_0 = 2.8$~fs.
    From~\cite{Ciappina_2015_PRL_114_143902}.
  }
\end{figure}

Envelope and carrier-wave Rabi flopping regimes are compared in Fig.~\ref{f:RabiFlopping}.
Here, the Bloch vector orbits around the equatorial plane with the transition frequency and slowly rotates between the south and north poles with the Rabi frequency.
The population inversion $w(t)$ exhibits a superposition of one slow oscillation $\sim \cos\tW{R}$ with rapid and weak \emph{Bloch--Siegert oscillations}~\cite{Bloch_1940_PR_57_522, Yan_2015_PRA_91_053834}.
The latter ones are small corrections from the counterrotating terms, which can be considered via the adiabatic perturbation theory~\cite{Ostrovsky_2004_PRA_70_033413}.

When the Rabi frequency is comparable to or larger than the laser frequency, $\g{RF} \gtrsim 1$, the contribution of counterrotating terms is more prominent, and the trajectory of the Bloch vector becomes sophisticated [Fig.~\ref{f:RabiFlopping}(b)].
The population inversion $w(t)$ now oscillates with the frequency comparable to that of the carrier wave and does not return to the initial value for $2\pi n$-pulses.
Thus the envelope area theorem no longer applies~\cite{Hughes_1998_PRL_81_3363}.

The resonant effects discussed also take place in periodic potentials.
In the regime defined by conditions $\g{RF} \gtrsim 1$ and $ \g{NP} = \Up / (\hbar\w0) \gtrsim 1$, one observes a complex interplay between intraband motion and phenomena where quantum coherence plays an essential role, such as Rabi oscillations~\cite{Wismer_2016_PRL_116_197401}.
Using the TDSE in the Houston basis, we define a generalized Rabi frequency applicable for periodic potentials and short pulses, which includes time dependence of the optical matrix element $\vec \xi_\mathrm{cv}\brK{t}$ and instantaneous detuning $\Delta\brK{t} = \E{cv}\brK{t}/\hbar - \w0$ due to field-induced intraband motion:
\begin{equation}\label{e:WRt}
  \Omega_\mathrm{R}(t) = \sqrt{\w{R}^2(t) + \Delta^2\brK{t}},
\end{equation}
where $\w{R}(t) = |e \vec F(t)\cdot \vec\xi_\mathrm{cv}\brK{t}|$.
The generalized pulse area can now be defined as
\begin{equation}\label{e:GA}
  \mathcal{A} = \int_{-\infty}^{\infty} \Omega_\mathrm{R}(t)\,\dd t.
\end{equation}

Numerical calculations~\cite{Wismer_2016_PRL_116_197401} show that the condition $\mathcal{A} = 2\pi n$ approximately coincides with the completion of an integer number of Rabi cycles at the $\Gamma$ point as long as $|\Delta\brK{t}| \lesssim \w{R}(t)$.
For higher fields, $\mathcal{A} / (2\pi)$ no longer counts Rabi cycles even if transitions to higher bands are neglected.
Near the resonance ($\hbar\w0 \approx \E{g}$), the cycle-averaged detuning is given by $\overline{\Delta\brK{t}} \approx [\overline{\E{cv}\brK{t}} - \E{g}] / \hbar = \Up / \hbar$.
Thus the applicability limit for the generalized Rabi frequency~\eqref{e:WRt} and pulse area~\eqref{e:GA} can be estimated by the following condition (see Table~\ref{t:gamma}):
\begin{equation}
  \g{RP} = \frac{\hbar\w{R}}{\Up} > 1.
\end{equation}

The last adiabaticity parameter in Table~\ref{t:gamma} that classifies field-matter interactions is $\g{RB}$, constructed as a ratio of the peak Rabi $\w{R}$ and Bloch $\w{B}$ frequencies.
This parameter describes an interplay between field-driven interband and intraband dynamics.
If interband and intraband transitions are induced by the same field, this is simply the ratio of the peak absolute value of the interband matrix element to the lattice constant:
\begin{equation*}
  \g{RB} = \frac{\w{R}}{\w{B}} = \frac{|\vec\xi_\mathrm{cv}^\mathrm{(max)}|}{a},
\end{equation*}
and thus this is the only parameter in Table~\ref{t:gamma} that depends only on material properties.

\begin{table}[t]
  \caption{\label{t:materials}
    A few representative materials commonly studied by modern ultrafast spectroscopy and their parameters.
    The absolute values of matrix elements $\bigl|\vec \xi_\mathrm{c,lh}^\mathrm{(max)}\bigr|$ are calculated using the VASP code~\cite{Kresse_1996_PRB_54_11169} with the TB09 meta-GGA functional~\cite{Tran_2009_PRL_102_226401} for a transition from the light hole band (lh) to the first conduction band (c).
    Band gaps and lattice constants are given according to~\textcite{LandoltBornstein_2002} for a temperature of $T = 300$~K.
    Crystal structures are denoted as follows: ``zb'' is zincblende, ``wz'' is wurtzite, ``fcc'' is face-centered cubic, and ``trig'' is trigonal.
  }\vspace{6pt}
  \begin{tabular}{ccccc}
    \hline\hline
    Material & Structure & $\E{g}$ (eV) & $a, c$ (\AA) & $\bigl|\vec \xi_\mathrm{c,lh}^\mathrm{(max)}\bigr|$ (\AA) \\ \hline
    GaAs             & zb   & 1.43 & 5.65       & 3.42 \\
    $\alpha$-GaN     & wz   & 3.45 & 3.19, 5.19 & 1.74 \\
    ZnO              & wz   & 3.3  & 3.26, 5.22 & 1.46 \\
    C (diamond)      & fcc  & 7.4  & 3.57       & 1.06 \\
    MgO              & fcc  & 7.8  & 4.2        & 0.96 \\
    $\alpha$-SiO$_2$ & trig & 9    & 4.9, 5.4   & 0.37 \\
    \hline\hline
  \end{tabular}
  \vspace{-6pt}
\end{table}

The values of $\g{RB}$ for a few representative solids can be obtained from the data in Table~\ref{t:materials}.
For the wide band gap dielectrics, the largest value of the interband matrix element can be much smaller than the lattice constant, while they are very close in semiconductors.

Materials with smaller band gap and effective masses tend to have larger $\g{RB}$, as predicted by Eq.~\eqref{e:xicv}, but remarkably all of them have $\g{RB} < 1$.
This condition guarantees that the electron traverses a large fraction of the Brillouin zone on a timescale that is short compared to that of interband transitions ($\w{R} < \w{B}$), which is a necessary condition for the applicability of single-band models.
For $\g{RB} \gtrsim 1$, one may expect an unusually strong dynamic Stark effect and inapplicability of results derived in the approximation of isolated bands, including the Houston and Kane states.
To the best of our knowledge, this limit does not exist in natural solids, but it might be realized in artificial periodic potentials and thus presents an interesting topic for future research.

\begin{figure}[t]
  \includegraphics[width=\linewidth]{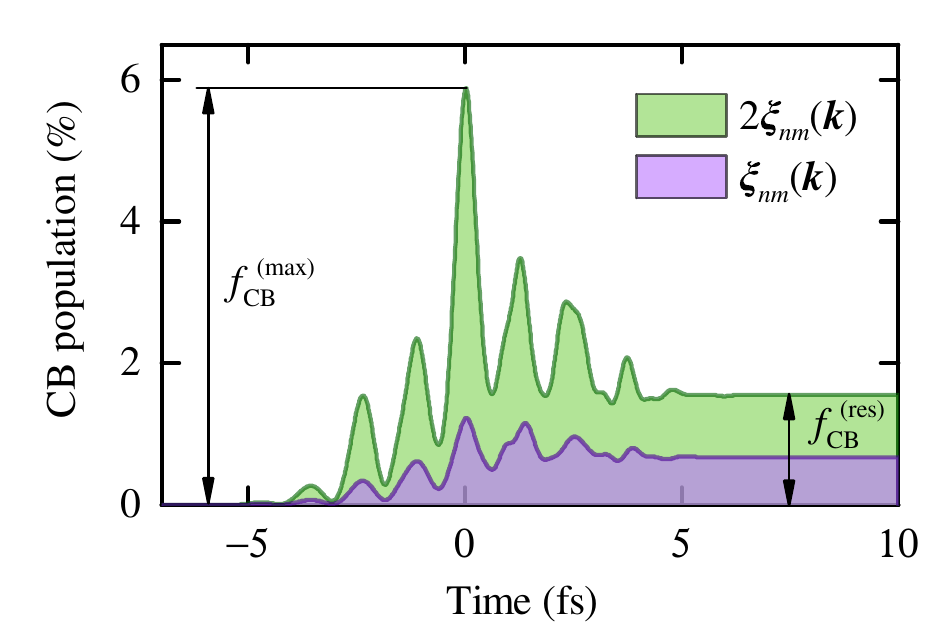}
  \caption{\label{f:ExcProb}%
    (Color online) Total population of six conduction bands for a few-cycle laser pulse with a sine-square envelope ($\lambda_0 = 750$~nm, $F_0 = 1$~V/\AA{}, FWHM = 3.7~fs).
    As optical matrix elements, we used those calculated with the meta-GGA TB09 functional (lower curve) and their artificially doubled values (upper curve).
  }
\end{figure}

Near a resonance, $\g{RB}$ counts the number of Rabi cycles per one Bloch period.
To illustrate the physical meaning of $\g{RB}$ far from a resonance, we compare in Fig.~\ref{f:ExcProb} the time-dependent excitation probability evaluated for $\alpha$-SiO$_2$ with that where we artificially doubled the amplitudes of optical matrix elements.
As one can see from Fig.~\ref{f:ExcProb}, the simulation with the doubled matrix elements shows an overall increase of the CB population, as well as an increase of the ratio of the peak transient population to the residual one $f_\mathrm{CB}^\mathrm{(max)}/f_\mathrm{CB}^\mathrm{(res)}$ from 1.8 to 3.8.
An increased amplitude of the matrix elements $|\xi_{nm}(\vec k)|$ yields larger nondiagonal terms of the field-matter interaction $\hat V_{nm}(t) = e \vec F(t) \cdot \vec \xi_{nm}\brK{t}$, which results in higher population transfer during and after the laser pulse.
This can be understood as a consequence of larger anticrossing gaps of the Wannier--Stark ladders in the nonresonant case (Fig.~\ref{f:WS-Kane}) or in terms of larger dynamic Stark shifts near the resonance (Fig.~\ref{f:DynStark}).

These observations suggest that materials with high amplitudes of optical matrix elements ($\g{RB} \lesssim 1$), e.g. III-V and nitride-based semiconductors, enter the tunneling regime at lower field intensities than insulators, where $\g{RB} \ll 1$.
From this point of view, the semiconductors might be more promising for potential applications relying on reversibility of the light-matter interaction~\cite{Schultze_2013_Nature_493_75, Novelli_2013_SR_3_1227, Lucchini_2016_Science_353_916, Sommer_2016_Nature_534_86}.
On the other hand, studying these phenomena in solids with smaller band gaps requires laser pulses in the mid-IR and THz domains.

Let us analyze how the adiabaticity parameters describing the nonresonant strong-field processes relate to the ones that characterize the near-resonant interaction.
Apart from obvious $N \approx 1$, $\g{NP}$ approaches $4\g{K}^{-2}$ because $\hbar\w0 \approx \E{g}$.
Using the explicit formula for the interband matrix element~\eqref{e:xicv}, one can express $\g{K}$ via the peak Rabi frequency:
\begin{equation}\label{e:gKres}
  \g{K} = \frac{\w0}{e F_0} \sqrt{\frac{m \E{g}}{1 + \beta^2}}
= \frac{\w0}{2\w{R}\sqrt{1 + \beta^2}} = \frac{1}{2 \g{RF}^{(0)}\sqrt{1 + \beta^2}}.
\end{equation}

This relation confirms that the adiabatic perturbation theory is dual to the conventional one employing the field-free basis, and the corresponding small parameters are inversely proportional to each other~\cite{Frasca_1998_PRA_58_3439}.

The dimensionless adiabaticity parameters introduced so far are summarized in Table~\ref{t:gamma}.
As one can see, all the field--matter interaction regimes discussed can be viewed from a single perspective as adiabatic or diabatic limits of one characteristic oscillatory motion with respect to another.
Our classification takes into account the latest developments in ultrafast laser spectroscopy, summarizes the well-established results, and indicates opportunities for further experimental and theoretical studies.

\subsection{Coherent dynamics and relaxation}

In the previous sections, we focused on perfectly coherent oscillatory processes.
However, this is an abstraction because all quantum-mechanical systems are connected to a dissipative environment and experience an irreversible decay of state population and quantum coherence.
These processes define another set of regimes, which are discussed next.

The timescales characterizing various scattering mechanisms in the electronic subsystem of solids excited by laser pulses cover more than 9 orders of magnitude: from nanosecond interband relaxation due to spontaneous photoemission to carrier-carrier scattering occurring on attosecond timescales.
According to~\textcite{Shah_1999}, the carrier relaxation processes can be classified into four temporally overlapping regimes, which are summarized in Table~\ref{t:4regimes} together with some typical processes.
The time scale for each event strongly depends on other parameters such as band structure, lattice temperature, carrier density, etc.

\renewcommand{\arraystretch}{1.12}
\begin{table}[!t]
\caption{\label{t:4regimes}%
Four relaxation regimes in photoexcited semiconductors and some typical processes~\cite{Shah_1999}.}
\begin{tabular}{p{\linewidth}}
\hline\hline
\centering\textit{Coherent regime} ($\lesssim 200$~fs)\tabularnewline\hline
Carrier-carrier scattering\\
Momentum scattering\\
Intervalley scattering ($\Gamma\rightarrow L, X$)\\
Hole-optical-phonon scattering\\
\hline\centering\textit{Non-thermal regime} ($\lesssim 2$~ps)\tabularnewline\hline
Electron-hole scattering\\
Electron-optical-phonon scattering\\
Intervalley scattering ($L, X \rightarrow\Gamma$)\\
Carrier capture in quantum wells\\
Intersubband scattering ($\Delta E > \hbar\w{LO}$)\\
\hline\centering\textit{Hot-excitation regime} ($\sim 1$--100~ps)\tabularnewline\hline
Hot-carrier-phonon interactions\\
Decay of optical phonons\\
Carrier-acoustic-phonon scattering\\
Intersubband scattering ($\Delta E < \hbar\w{LO}$)\\
\hline\centering\textit{Isothermal regime} ($\gtrsim 100$~ps)\tabularnewline\hline
Carrier recombination\\
\hline\hline
\end{tabular}
\end{table}

Relaxation processes strongly influence optical properties of solids and place requirements on temporal characteristics of laser pulses for observing a particular ultrafast phenomenon.
The main requirement here is that some characteristic time of light-matter interaction $T$ should be much smaller than the population relaxation time $T_1$ and, for phase-sensitive processes, it should also be smaller than the phase relaxation time $T_2$.
These conditions can be fulfilled by using shorter and stronger laser pulses.
It is also possible to study the same physical phenomena in artificial periodic structure with much longer relaxation times than those in natural solids, e.g., $T_1 \sim 100$~ms in optical lattices.

Currently, modeling of ultrafast phenomena is based on the numerical solution of the TDSE~\cite{Bachau_2006_PRB_74_235215, Korbman_2013_NJP_15_013006, Wu_2015_PRA_91_043839}, density-matrix method~\cite{Lindberg_1988_PRB_38_3342, HaugKoch_2009} and time-dependent Kohn--Sham equations~\cite{Yabana_2012_PRB_85_045134, Andrade_2015_PCCP_17_31371, Otobe_2016_PRB_93_045124, Tancogne-Dejean_2017_PRL_118_087403}.
In strong laser fields, charge carriers are often treated as independent (quasi)particles, and their interaction with the environment is neglected.
The assumption of purely coherent dynamics on a few-fs timescale was motivated by previous experimental studies of relaxation phenomena in semiconductors, where the measured dephasing time $T_2$ varies from tens to hundreds of femtoseconds~\cite{Oudar_1985_PRL_55_2074, Becker_1988_PRL_61_1647, Prabhu_1997_APL_70_2419}.
Laser pulses of a much shorter duration are routinely available nowadays, so the phase relaxation is not expected to significantly influence coherent dynamics during a time interval much shorter than $T_2$.

Nevertheless, comparison of simulations with the recent experimental data on high-harmonic generation in solids~\cite{Ghimire_2011_NP_7_138, Luu_2015_Nature_521_498, Schubert_2014_NP_8_119, Vampa_2014_PRL_113_073901} demonstrated the significance of dephasing in the presence of a strong field.
Remarkably, this phenomenon was also demonstrated experimentally in 2D materials~\cite{Liu_2017_NP_13_262, Cox_2017_NC_8_14380}.
These results suggest that interband polarization is overestimated in the independent-particle approximation, and decoherence phenomena are still important on a few-fs timescale.

The problem with overestimated interband coherencies was addressed phenomenologically in the density-matrix models, where dephasing is introduced in the Markov approximation with $T_2$ as a free parameter~\cite{Golde_2008_PRB_77_075330, Schubert_2014_NP_8_119, Higuchi_2014_PRL_113_213901, Pati_2015_NC_6_7746, Luu_2015_Nature_521_498, Vampa_2014_PRL_113_073901, Langer_2017_NP_533_225}.
Reaching an agreement with experimental data required very short dephasing times $T_2 \sim 1$--3~fs.
Intense laser pulses drive the charge carriers within a large part of the BZ and induce a highly nonequilibrium population distribution, which may extend up to several conduction and valence bands.
In such conditions, the probability of various scattering processes can be much higher than that in weaker fields.

\begin{figure}[!ht]
  \includegraphics[width=\linewidth]{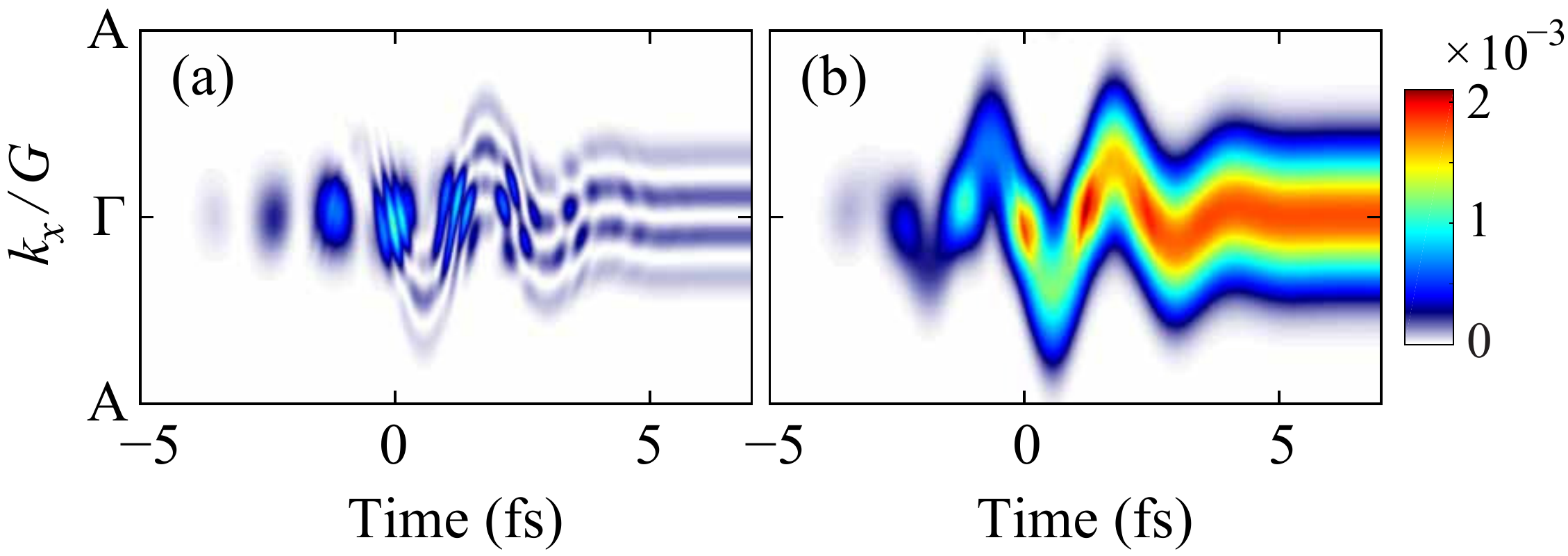}
  \caption{\label{f:Dephasing}
    (Color online) Simulations of population evolution in the BZ of wurtzite GaN (a) without dephasing and (b) for $T_2 = 2$~fs.
    Field amplitude is $F_0 = 0.5$~V/\AA{}, and other pulse parameters are the same as in Fig.~\ref{f:ExcProb}.
}
\end{figure}

To illustrate the physical consequences of dephasing, we compared two quantum-mechanical simulations.
Figure~\ref{f:Dephasing}(a) shows the evolution of the population induced by the laser field interacting with a crystal in the model without dephasing.
It features the field-resolved oscillations and multiple interference fringes on the population distribution during and after the pulse.
Figure~\ref{f:Dephasing}(b), calculated with $T_2 = 2$~fs, shows a single Gaussian-like wavepacket moving according to the acceleration theorem and predicts a much higher residual population.
Thus the ultrafast decay of interband coherencies makes the quantum evolution of electrons closer to the semiclassical model based on Boltzmann equations, where nondiagonal density-matrix elements are neglected and interband transitions are described by rates of population change~\cite{Jacoboni_2010, Rossi_2011}.

The ultrafast scattering times as short as a few femtoseconds previously appeared in theoretical treatments of high-field transport in dielectrics~\cite{Fischetti_1987_PRB_35_4404, Arnold_1994_PRB_49_10278} and electron-hole plasma in optically-excited semiconductors~\cite{Binder_1992_PRB_45_1107, Scott_1992_PRL_69_347}.
These results were inconsistent with the experimentally measured values of $T_2 \sim 20$--100~fs~\cite{Oudar_1985_PRL_55_2074, Prabhu_1997_APL_70_2419}.
For that problem, the Markov approximation was identified as the main source of the unexpectedly fast dephasing.

It is well known that both Boltzmann's \emph{Sto\ss{}zahl} \emph{ansatz} and the Markov approximation rely on the existence of two well-separated time scales: a slow time scale of the system and a fast time scale characterizing the decay of bath correlation functions~\cite{Carmichael_2013}.
In other words, all collision processes are considered to be pointlike and instantaneous: the mean scattering time $\tau_\mathrm{s}$ between two successive collisions should be much longer than the collision duration $\tau_\mathrm{c}$~\cite{Rossi_2011}.
Obviously, this condition is not satisfied in modern experiments with few-cycle laser pulses, where charge carriers can be controlled on the attosecond timescale.
Another scattering process may start before the completion of a previous one, which cannot be adequately described within the completed-collision and Markov approximations.

The problem of unphysically fast dephasing was previously solved by employing more advanced quantum-kinetic models describing screened carrier-carrier scattering beyond the Markov approximation~\cite{Hohenester_1997_PRB_56_13177, Banyai_1998_PRL_81_882, Kremp_1999_PRE_60_4725, Gartner_2000_PRB_62_7116, HaugJauho_2008, Bonitz_2016} and taking into account the finite time required for the formation of screening ``clouds'' around individual particles.
This phenomenon was confirmed later by experimental observations~\cite{Huber_2001_Nature_414_286}, and it is neglected in the Markov approximation, where all scattering events are treated as instantaneous.

The major role of electron-electron interaction in the decay of interband coherence is supported by a recent comparison of experimental data to simulations of HHG in SiO$_2$ with semiconductor Bloch equations including the Hartree--Fock terms~\cite{Garg_2016_Nature_538_7625}.
It was shown that interband polarization in the presence of an interaction became significantly smaller than that in the independent-particle approximation.
Nevertheless, it is still an open question if the ultrafast dephasing is physically meaningful for solids exposed to intense few-cycle pulses and which quantum-kinetic model gives the best compromise between the computational complexity and completeness of theoretical description.

\section{Strong-field phenomena and modern ultrafast spectroscopy}
\label{s:3}

In this section, we discuss a few recently investigated strong-field phenomena in solids.
Wherever appropriate, we point out differences to similar studies on artificial periodic structures.
In the final part, we discuss the phenomena related to a geometric phase and perspectives for their investigation in bulk and low-dimensional materials with the modern methods of ultrafast laser spectroscopy.

\subsection{Bloch--Zener oscillations and high-harmonic generation}

Already in the early work on band theory,~\textcite{Bloch_1929_ZP_52_555} showed that, in the absence of scattering and interband transitions, the crystal momentum $\vec K(t)$ of an electron in a constant electric field is a linear function of time.
The infinite growth of $\vec K(t)$, which also follows from the acceleration theorem~\eqref{e:Bloch_k}, is a feature of the extended zone scheme.
In the reduced zone scheme, the crystal momentum oscillates with a frequency of $\w{B} = |e F_0| a$.
Each time an electron wavepacket crosses the BZ boundary, its group velocity changes its sign in agreement with Eq.~\eqref{e:Bloch_r} and the periodicity of band dispersion.
These predictions were not confirmed by experimental observations for several decades because Bloch oscillations can be detected only if a sufficient number of Bloch cycles $T_\mathrm{B} = 2\pi\hbar/(|e F_0| a)$ is completed within the momentum relaxation time $T_1^\mathrm{intra} \sim 100$~fs, and the Bloch frequency exceeds the rate of interband transitions.
These conditions were first realized experimentally in artificial superlattices~\cite{Esaki_1970_IJRD_14_61}, which consist of alternating layers of semiconductors with different band gaps.
The periodicity of potential in the growth direction is given by the sum $d = a + b$ of layer thicknesses $a$ and $b$.
Typical values of $d$ are from a few to tens of nanometers, which is by 1 or 2 orders of magnitude larger than the lattice constant of bulk crystals.
This additional periodicity modifies the energy spectrum and creates minibands.
The corresponding Brillouin minizone size $2\pi/d$ is much smaller than that in a bulk semiconductor, which allows the condition $T_\mathrm{B} \ll T_1^\mathrm{intra}$ to be satisfied at field amplitudes much lower than necessary in bulk solids.
Therefore, in superlattices, charge carriers more easily reach the upper part of the conduction band before being incoherently scattered, which makes these quasi-one-dimensional structures ideal for studying various strong-field phenomena~\cite{Ivchenko_1997, Leo_2003, Tsu_2011}.

In addition to various scattering processes, the periodic motion of carriers is impeded by interband transitions.
The simple models considering only one or two isolated bands within the nearest-neighbor tight-binding model usually predict Wannier--Stark localization and band collapse in the strong-field limit~\cite{Wannier_1937_PR_52_191, Dunlap_1986_PRB_34_3625, Holthaus_1992_PRL_69_351, Gruzdev_2007_PRB_75_205106, Apalkov_2012_PRB_86_165118}.
However, numerical simulations with a sufficient number of bands, as well as electro-optical measurements in superlattices demonstrate splitting and delocalization of electrons in high fields due to Zener breakdown~\cite{Sibille_1998_PRL_80_4506, Rosam_2001_PRL_86_1307, Glutsch_1999_PRB_60_16584, Breid_2006_NJP_8_110, Dreisow_2009_PRL_102_076802, Longhi_2012_JPB_45_225504}.
Therefore, in a realistic system, there is always an interplay between Bloch oscillations and tunneling to other bands---when a wavepacket crosses the Brillouin-zone boundary, it splits into two parts.
In the literature, this phenomenon is known as the \emph{Bloch--Zener oscillations}.

\begin{figure}[t]
  \includegraphics{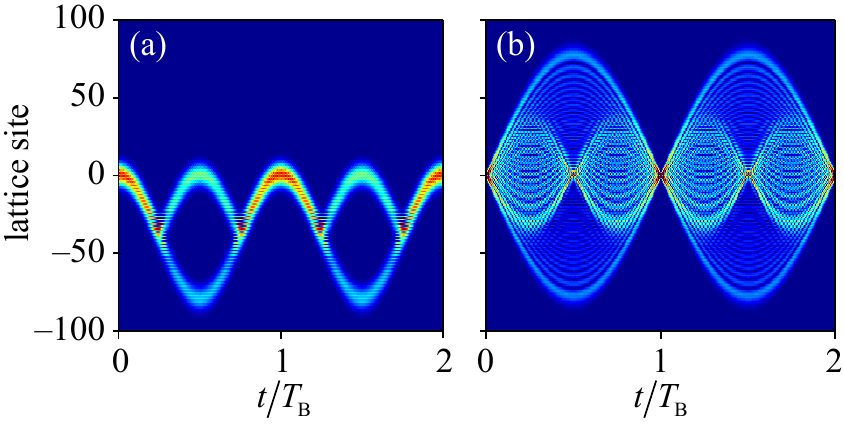}
  \caption{\label{f:BlochZener}%
    (Color online)
    Dynamics of an oscillating and breathing Bloch--Zener modes for spatially localized wave packets exposed to a constant electric field (two-band model).
    The probability density $|\psi(\vec r, t)|^2$ of a double-periodic lattice plotted vs coordinate and time.
    Oscillations with a smaller period take place due to the intraband motion only in the lowest band, while those with a larger period are due to Bragg reflections after tunneling to the second band.
    (a)~The initial state is a broad Gaussian distribution in real space (the FWHM is about 10 lattice sites).
    The plot demonstrates the Bloch oscillations of translational motion and splitting of the wavepacket in real space, which occurs as the wavepacket crosses the BZ edges in reciprocal space (see Fig.~\ref{f:Bands}).
    (b)~The wavepacket is initially located at a single lattice site and demonstrates oscillations of its width.
    From~\textcite{Breid_2006_NJP_8_110}.
  }
\end{figure}

As already mentioned in Sec.~\ref{s:BO}, Bloch oscillations limit the translational motion of an electron wavepacket and its spatial extent.
These localization dynamics are known as oscillating and breathing modes, respectively.
Figure~\ref{f:BlochZener} shows that both of these modes are in interplay with interband transitions (Zener tunneling).
This interplay splits the real-space density distribution and creates interference fringes.

Recent progress in the generation of ultrashort laser pulses has enabled experimental investigation of dynamic Bloch oscillations in bulk solids via high-order harmonics emitted by highly nonlinear field-induced intraband currents~\cite{Ghimire_2011_NP_7_138, Schubert_2014_NP_8_119, Luu_2015_Nature_521_498, Garg_2016_Nature_538_7625, You_2017_NP_13_345}.
Remarkably, the measurements in solids have demonstrated much higher influence of distant neighbors (up to $\ellmax = 6$) than in superlattices.
This can be explained by the fact that the lattice period in solids ($a \sim 5$~\AA{}) is smaller by 1 or 2 orders of magnitude than that in superlattices ($d \sim 1$--10~nm), while the coherence length is nearly the same.

HHG in atoms was understood within a simple semiclassical three-step model including ionization, acceleration, and recollision of electrons with the original atom~\cite{Corkum_1994_PRL_71_1994, Lewenstein_1994_PRA_49_2117}.
The physics of HHG in solids is more sophisticated, and its description requires a comprehensive quantum-mechanical modeling of electron dynamics in a periodic potential.
The generation of harmonics in solids can be attributed to both interband and intraband components of the total current, while their relative contribution strongly depends on material, driving pulse parameters, and the spectral range where harmonics are observed.
Currently, the following mechanisms of HHG are commonly discussed: %
a generalization of the three-step model considering the electron-hole recollision in the real- and reciprocal-space pictures~\cite{Higuchi_2014_PRL_113_213901, Osika_2017_PRX_7_021017, Vampa_2015_PRL_115_193603},
direct interband transitions and their interference due to the presence of multiple valence and conduction bands~\cite{Hawkins_2015_PRA_91_013405, Hohenleutner_2015_Nature_523_572, Wu_2016_PRA_94_063403, Du_2017_OE_25_151},
and deceleration of carrier's intraband motion due to Bragg reflections on the lattice potential~\cite{Feise_1999_APL_75_3536, Ghimire_2011_NP_7_138, Luu_2015_Nature_521_498, Luu_2016_PRB_94_115164, Muecke_2011_PRB_84_081202}.

Both interband and intraband components may feature a linear scaling of the cutoff frequency with the field~\cite{Ghimire_2011_NP_7_138, Luu_2015_Nature_521_498, Hohenleutner_2015_Nature_523_572, Vampa_2015_Nature_522_462}, which hampers unambiguous identification of the dominant contribution and requires further analysis as well as comparison of experimental data with rigorous numerical simulations.
In the recent publications, the following additional characteristics of HHG radiation have been discussed: %
frequency dependence of the group delay~\cite{Wu_2016_PRA_94_063403, Hohenleutner_2015_Nature_523_572, Garg_2016_Nature_538_7625}
and the scaling of individual harmonics peak intensity with the driving field amplitude~\cite{Ghimire_2011_NP_7_138, Schubert_2014_NP_8_119, Luu_2015_Nature_521_498, Hohenleutner_2015_Nature_523_572}.

High-harmonic spectroscopy presents a valuable alternative to scanning tunneling microscopy (STM), electron diffraction, and angle-resolved photoemission spectroscopy (ARPES).
Because of the high sensitivity of high-order harmonics to crystallographic orientations~\cite{Ghimire_2011_NP_7_138, Luu_2015_Nature_521_498, You_2017_NP_13_345, Langer_2017_NP_533_225} and the joint density of states~\cite{Tancogne-Dejean_2017_PRL_118_087403}, the electronic band structure and lattice potential can be reconstructed from the high-harmonic spectra~\cite{Luu_2015_Nature_521_498, Vampa_2015_PRL_115_193603}.
Moreover, it offers the ability to explore dynamic changes of this structure under the influence of strong fields.

\subsection{Light-waveform control of electric current}

Even though the motion of an electron driven by a laser pulse is determined by the electric field, experimental observables frequently depend only on cycle-averaged quantities.
In this case, the envelope and the instantaneous frequency of the pulse fully determine measurement outcomes.
The development of experimental techniques sensitive to the carrier--envelope phase (and hence the waveform) of laser pulses in the visible and infrared spectral range marked an important milestone in ultrafast optics and served as a basis for attosecond science~\cite{Brabec_2000_RMP_72_545}.
Light-waveform control of phase-sensitive processes consists of driving them with controlled optical fields.
In atomic and molecular physics, light-waveform control of electron motion has revolutionized time-resolved measurements~\cite{Krausz_2009_RMP_81_163}.
This type of control in solids is a relatively new development, but it may have a similar long-term impact on ultrafast metrology and spectroscopy~\cite{Krausz_2014_NP_8_205}.

A precursor of this development was coherent control of photocurrents in molecular chains and semiconductors~\cite{Kurizki_1989_PRB_39_3435, Hache_1997_PRL_78_306, Atanasov_1996_PRL_76_1703, Rioux_2012_PE_45_1, Shapiro_2012}.
It was found that a current can be induced in an unbiased semiconductor by exciting charge carriers with two pulses that have different frequencies, while the direction of the current depends on their relative phase.
Typically, the pulses would have central frequencies $\omega$ and $2 \omega$, and the second harmonic would be resonant with the transition between the upper valence and the lower conduction bands of the semiconductor.
The interference between one- and two-photon transitions breaks the symmetry of interband excitations, so the band population at a crystal momentum $\vec k$ may be different from that at $-\vec k$.
A similar kind of interference has also been observed with carrier-envelope-phase- (CEP-) stabilized laser pulses that had a spectral width exceeding an optical octave, so that components at frequencies $\omega$ and $2\omega$ could be found within a single pulse~\cite{Fortier_2004_PRL_92_147403}.

The interference between one- and two-photon transitions induced by relatively weak fields has a straightforward generalization for shorter and stronger nonresonant pulses ($\g{K} \gtrsim 1$) as the interference between $n$- and $m$-photon transitions driven by CEP-stabilized pulses ($n$ and $m$ are integer numbers of opposite parity).
Multiphoton transitions require a high intensity of the laser pulse, but if its central frequency is well below the absorption edge, the pulse spectrum can be contained within the transparency window and therefore the solid can withstand the high intensity.
The interference of multiphoton excitation pathways as a mechanism of CEP control was proposed by~\textcite{Kruchinin_2013_PRB_87_115201} and further developed experimentally and theoretically in~\textcite{Paasch-Colberg_2016_Optica_3_1358}.

At yet higher intensities ($\g{K} \lesssim 1$), the transition from the multiphoton to the tunneling regime facilitates the sensitivity of the excitation rate to the pulse waveform.
Light-field control of a current in this regime was first demonstrated in a dielectric (SiO$_2$) exposed to few-cycle near-infrared linearly polarized laser pulses~\cite{Schiffrin_2013_Nature_493_70}.
The dielectric sample was placed between two electrodes, and even though no bias was applied, the circuit connecting the electrodes detected that sufficiently intense laser pulses induced a charge displacement equal to the absolute value of a total residual polarization including both interband and intraband contributions:
\begin{equation*}
  Q = |\vec P(t\rightarrow +\infty)| = S \int_{-\infty}^{+\infty} \vec J(t)\,\dd t.
\end{equation*}
Here $S$ is an effective surface area that is perpendicular to the total current density $\vec J(t)$.

\begin{figure}[t]
  \includegraphics[width=\linewidth]{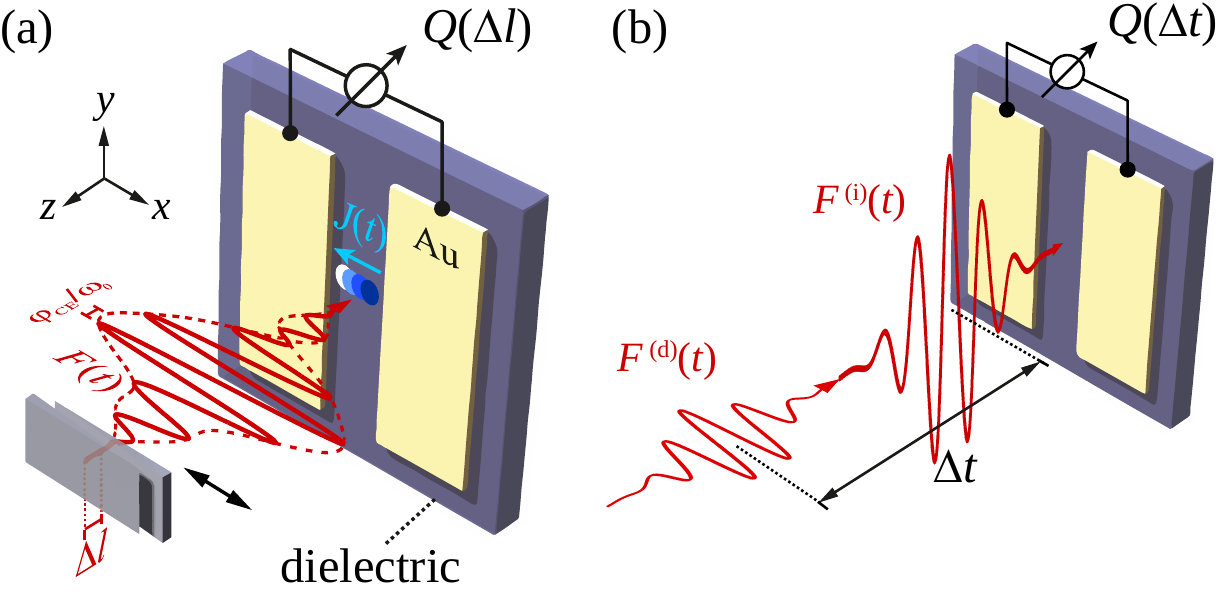}
  \caption{\label{f:WaveformControl}%
    (Color online) Schematic illustration of light-waveform control experiments in a solid.
    (a)~Single-pulse arrangement.
    A dielectric surface patterned with gold electrodes was exposed to a CEP-controlled few-cycle IR pulse $F(t)$.
    The pulse CEP was changed by varying propagation length $\Delta l$ in the fused-silica wedges.
    (b)~Injection-drive arrangement.
    Two orthogonally polarized IR laser pulses, delayed by $\Delta t$, irradiate a metal-dielectric-metal junction.
    The strong injection pulse with $F^\mathrm{(i)} \sim 1$~V/\AA{} excites charge carriers in a dielectric, and the weak drive field $F^\mathrm{(d)} \sim 0.05$~V/\AA{} displaced them towards the electrodes.
  }
\end{figure}

Experiments with isolated CEP-stabilized laser pulses [Fig.~\ref{f:WaveformControl}(a)] showed that the residual polarization can be controlled by the carrier-envelope phase.
To obtain temporal resolution, these measurements were also performed in a pump-probe fashion, where an intense laser pulse $F^\mathrm{(i)}$ polarized along the electrodes injected electrons from valence to conduction bands, while an orthogonally polarized weak drive pulse $F^\mathrm{(d)}$ displaced the excited carriers toward the electrodes  [Fig.~\ref{f:WaveformControl}(b)].
The dependence of the transferred charge on the delay resembled the drive waveforms, providing evidence for light-field control and suggesting that the charge-carrier injection occurs on a $\sim 1$~fs timescale.

The physical mechanism leading to the light-waveform controlled current has been a matter of debate.
The proposed interpretations engage the dynamic formation of Wannier--Stark states in the adiabatic tunneling limit~\cite{Schiffrin_2013_Nature_493_70, Kwon_2016_SR_6_21272}, the interference of excitation channels in the multiphoton and tunneling regimes~\cite{Kruchinin_2013_PRB_87_115201, Paasch-Colberg_2016_Optica_3_1358}, the field-driven intraband motion of charge carriers after their excitation~\cite{Foldi_2013_NJP_15_063019, Yakovlev_2016}, and the dynamics of virtual electron-hole pairs~\cite{Yablonovitch_1989_PRL_63_976, Krausz_2014_NP_8_205, Khurgin_2016_JOSAB_33_C1}.
Models based on these concepts differ in the assumptions and representations that they use, so the physical insights that they give are particularly relevant in different regimes and limiting cases.
The status quo is that, in general, light-field controlled charge transfer emerges as a result of interplay between interband and intraband dynamics.

In comparison to the general concept of coherent control, light-waveform control of the electric current relies on a nearly adiabatic evolution of some quantity describing carrier dynamics with respect to the laser field.
The type and degree of adiabaticity can be analyzed using the parameters discussed in Sec.~\ref{s:2}.
For example, the Keldysh parameter $\g{K}$ determines the applicability of the quasistatic approximation to a carrier excitation rate.
The dynamic localization parameter is particularly important in this context:
for $\g{DL} \ll 1$, the applicability of the effective-mass approximation justifies the adiabaticity of instantaneous group velocity $\vec v_n\brK{t}$ with respect to the field waveform, while $\g{DL} \gtrsim 1$ calls for taking into account the dynamic Bloch oscillations (see Sec.~\ref{s:BO}).

In the pump-probe experiments, light-field control naturally emerges if the carrier excitation by an injection pulse occurs during a time interval much shorter than the period of a weak drive field.
In this case, the interband and intraband dynamics occur on different time scales and thus become adiabatically decoupled.
If the drive field is weak enough ($\g{DL} \ll 1$), the group velocity of carriers is given by Eq.~\eqref{e:vnEMA}, and the transferred charge can be written as a convolution of the vector potential of the drive field inside a solid $A^\mathrm{(d)}(t) = - \int_{-\infty}^{t} F^\mathrm{(d)}(t_1)\,\dd t_1$ with the time-dependent band population $f_n(\vec k, t)$~\cite{Paasch-Colberg_2016_Optica_3_1358}:
\begin{equation}\label{e:QEMA}
  Q^\mathrm{(EMA)}(\Delta t) \approx \sum_{n, \vec k} \frac{2e^2 S}{m_n} \int_{-\infty}^{+\infty} f_n(\vec k, t) A^\mathrm{(d)}(t - \Delta t)\,\dd t.
\end{equation}
Here $\Delta t$ is the delay between injection and drive pulses.

The deconvolution of Eq.~\eqref{e:QEMA} resolves the real-time waveform of the drive field in a crystal.
This expression also reveals a close analogy between the two-pulse measurement of $Q(\Delta t)$ in a dielectric and attosecond streaking in vacuum~\cite{Itatani_2002_PRL_88_173903, Kienberger_2004_Nature_427_817}.
However, in the limit of a strong drive field ($\ell \g{DL} \gg 1$), one does not have such a simple relation, and the deconvolution of $Q(\Delta t)$ will also require prior knowledge of carrier group velocities in the entire BZ.

Several recent publications have identified potential applications for waveform-controlled electric currents.
One of them is a solid-state device for measuring the carrier-envelope phase~\cite{Paasch-Colberg_2014_NP_8_214} and waveform~\cite{Schiffrin_2013_Nature_493_70, Paasch-Colberg_2016_Optica_3_1358} of few-cycle laser pulses.
This application exemplifies time-resolved measurements with a subfemtosecond sampling signal, which is referred to as attosecond metrology~\cite{Hentschel_2001_Nature_414_509, Krausz_2014_NP_8_205}.
Employing the waveform control to investigate and possibly extend the frontiers of signal processing in electronics is another intriguing idea~\cite{Krausz_2014_NP_8_205, Wachter_2015_NJP_17_123026, Lee_2016_PRL_116_057401}.
Even from a purely academic perspective, studying the nonequilibrium electron dynamics on the femtosecond scale present numerous research opportunities, one of which is the transition from ballistic to dissipative electron transport in a pump-probe setting~\cite{Wachter_2014_PRL_113_087401}.

\subsection{Dynamic Franz--Keldysh effect}

A constant electric field applied to a semiconductor or insulator changes the optical properties of the solid: the absorption edge shifts to smaller energies, and the spectrum above the band edge acquires an oscillatory behavior.
Theoretically, these effects can be described as field-induced renormalization of electronic wavefunctions in a tilted lattice potential.
Using Eqs.~\eqref{e:Kane} and~\eqref{e:eta}, one can show that, in the effective-mass approximation~\eqref{e:EnkEMA}, the component of hybrid Kane functions that is parallel to the field polarization is given by the Airy function, while the other two components remain the Bloch waves~\cite{Tharmalingam_1963_PR_130_2204}.
The electronic wavefunction of valence and conduction bands coupled by a static field can be written as
\begin{equation}
  \varphi(\vec k_\perp, \vec r) \propto \Ai\left(\frac{|eF_0| x - \epsilon}{\hbar\theta_x}\right) \e^{i (k_y y + k_z z)},
\end{equation}
\begin{gather*}
  \hbar\theta_x = \left[\frac{(e F_0 \hbar)^2}{2 m}\right]^{1/3}
,\;
  \epsilon = \mathcal{E} - \frac{\hbar^2(k_y^2 + k_z^2)}{2m}
.
\end{gather*}

\begin{figure}[!ht]
\includegraphics[width=\linewidth]{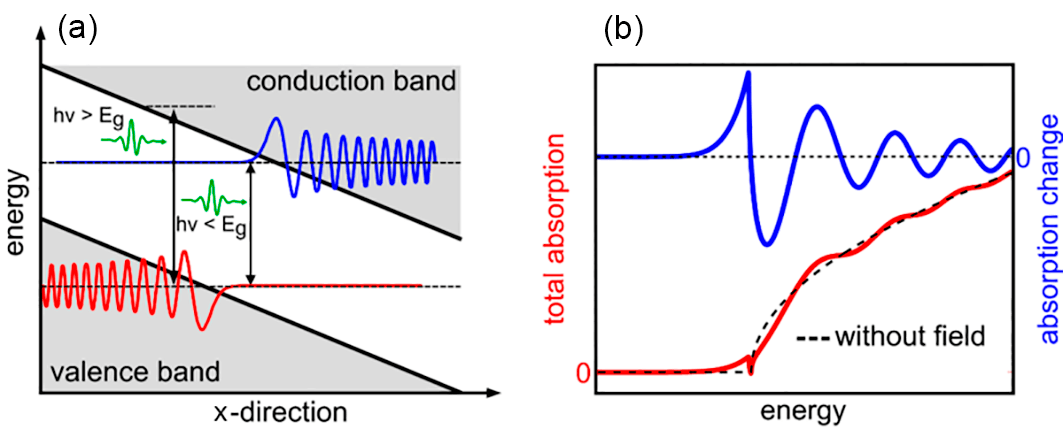}
\caption{\label{f:FranzKeldysh}%
  (Color online) Schematic illustration of the Franz--Keldysh effect.
  (a)~The wavefunctions of the valence and conduction band (lower red and upper blue curves, respectively) in a constant field are approximately given by the Airy function, which exponentially decays into the band gap and oscillates in the allowed regions.
  (b)~The absorption spectra of a bulk semiconductor with (lower curve) and without the external field (dashed black curve) and their difference $\Delta \alpha = \alpha(\omega, F_0) - \alpha(\omega, 0)$ (upper curve).
  Courtesy of A.~Leitenstorfer and Ch.~Schmidt, University of Konstanz.
}
\end{figure}

The Airy function features an exponential tail leaking inside the forbidden energy gap as well as decaying oscillations outside it [Fig.~\ref{f:FranzKeldysh}(a)].
The formation of these features has a profound effect on both real and imaginary parts of dielectric permittivity, which can be probed by a weak resonant field [Fig.~\ref{f:FranzKeldysh}(b)].
As a result, the \emph{photon-assisted tunneling} with absorption below the band gap $\hbar\omega < \E{g}$ becomes allowed,
and the field-dependent absorption coefficient can be approximated by the following expression~\cite{Franz_1958_ZNA_13_484, Keldysh_1958_JETP_34_788, Tharmalingam_1963_PR_130_2204}
\begin{equation}\label{e:awF0}
  \alpha(\omega, F_0) \propto \frac{\theta_x^{3/2}}{\E{g} - \hbar\omega}\exp\left[-\frac{4}{3} \left(\frac{\E{g} - \hbar\omega}{\hbar\theta_x}\right)^{3/2}\right]
\end{equation}

Promptly following the theoretical predictions by \cite{Franz_1958_ZNA_13_484} and \cite{Keldysh_1958_JETP_34_788}, the FKE was confirmed by experimental observations in bulk semiconductors~\cite{Boer_1958_NWS_45_460, Boer_1959_ZP_155_170}.
Further developments addressed contributions from multiple bands and band nonparabolicities~\cite{Aspnes_1974_PRB_10_4228, Hader_1997_PRB_55_6960}, excitonic effects~\cite{Ralph_1968_JPC_1_378, Rowe_1970_PRL_25_162, Dow_1970_PRB_1_3358, Blossey_1971_PRB_3_1382, DuqueGomez_2015_JPCS_76_138, Pedersen_2015_PLA_379_1785}, quantum confinement in semiconductor nanostructures~\cite{Miller_1986_PRB_33_6976, Hache_1989_APL_55_1504, Schmeller_1994_APL_64_330, Hughes_2000_PRL_84_4228}, phonon-assisted and multiphoton transitions~\cite{Penchina_1965_PR_138_A924, Hassan_1975_NCB_28_27, Wahlstrand_2011_PRB_83_233201},
and harmonically varying strong field~\cite{Jauho_1996_PRL_76_4576, HaugJauho_2008, Otobe_2016_PRB_93_045124}.

The experimental study of the FKE in static electric fields applied to semiconductors is restricted to field strengths on the order of $\sim 10^5$--$10^6$~V/cm due to Zener breakdown.
Dielectric permittivity changes little in these fields, which makes it difficult to observe the exponential tail and oscillations shown in Fig.~\ref{f:FranzKeldysh}(b).
The change in permittivity can be measured with high accuracy by applying an oscillating electric field and using a lock-in amplifier~\cite{Hamaguchi_2013}.
This method, known as \emph{electromodulation spectroscopy}, has been used since the 1960s to provide valuable information on the density of states and band structure of semiconductors~\cite{Frova_1966_PR_145_575, Aspnes_1972_PRL_28_168, Aspnes_1973_SS_37_418}.
Field strengths substantially exceeding $10^5$~V/cm were achieved by using short and intense laser pulses with a central photon energy much smaller than fundamental band gap~\cite{Nordstrom_1997_PSSB_204_52, Hughes_1999_PRB_59_R5288, Chin_2001_PRL_86_3292}.

An important parameter describing the Franz--Keldysh effect in a time-dependent pump field is the ratio of the ponderomotive energy $\Up$ to the pump photon energy, that is, $\g{NP}$.
The static FKE corresponds to $\g{NP} \gg 1$.
For $\g{NP} \ll 1$, the induced change of absorption is due to multiphoton processes.
The intermediate regime $\g{NP} \sim 1$ is known as the \emph{dynamic Franz--Keldysh effect} (DFKE)~\cite{Jauho_1996_PRL_76_4576, Nordstrom_1997_PSSB_204_52, Nordstrom_1998_PRL_81_457, HaugJauho_2008}.
The strong-field approximation~\cite{Keldysh_1965_JETP_20_1307, Reiss_1980_PRA_22_1786} discussed in Section~\ref{s:Keldysh} shows that, in this regime, the lowest-order multiphoton excitation channel is closing due to the increase of the effective band gap $\widetilde E_\mathrm{g} = \E{g} + \Up$, and electron wavefunctions start penetrating the classically forbidden regions.
Thus the diabatic tunneling regime is a minimal prerequisite for observing DFKE.

Modern studies of the DFKE, in both theory~\cite{Jauho_1996_PRL_76_4576, Nordstrom_1997_PSSB_204_52, Platero_2004_PR_395_1, Otobe_2016_PRB_93_045124} and experiment~\cite{Ghimire_2011_PRL_107_167407, Schultze_2013_Nature_493_75, Novelli_2013_SR_3_1227, Schultze_2014_Sci_346_1348, Lucchini_2016_Science_353_916}, are focused on the subcycle control of optical properties of semiconductors and dielectrics by employing intense THz or IR few-cycle pulses.
The waveform control of absorption is achieved closer to the adiabatic tunneling limit, where $\g{K} \ll 1$ and $ \g{NP} \gg 1$, and it usually requires higher pump fields.

The Franz--Keldysh effect is at the heart of modern electroabsorption modulators, which are able to operate at low voltage (only a few volts) and with modulation bandwidth up to tens of gigahertz~\cite{Lach_2005_JOFCR_2_140}.
These devices are widely used in optical fiber communications and integrated optoelectronics~\cite{Chuang_2009, Ebeling_2012}.
The experimentally demonstrated switching of optical absorption on the subfemtosecond timescale may pave the way toward the extension of light modulation and thereby optical signal processing from the gigahertz to the terahertz and petahertz regimes.

\subsection{Energy transfer between light and matter}

Employing strong-field phenomena for high-speed metrology and signal processing has a major challenge: irreversible energy transfer per switching cycle.
At the same time, transient energy exchange is a prerequisite for reversible nonlinear processes.
Therefore it is necessary to precisely measure and understand minuscule amounts of energy transferred between light and matter.
The energy flow is determined by the electric field $\vec F(t)$ of a light pulse and the polarization response $\vec P(t)$ of the medium:
\begin{equation}\label{e:dWdt}
  \frac{\dd W}{\dd t} = \vec F(t) \cdot \frac{\dd \vec P}{\dd t} = \vec F(t) \cdot \vec J(t).
\end{equation}
Here $W$ is work per unit volume.
We do not distinguish between bound and free charge carriers, so $\vec J(t)$ includes the displacement current.
For a sample so thin that $\vec F(t)$ changes little during propagation, $\vec J(t)$ can be reconstructed from the incident and transmitted fields provided that these fields are measured unambiguously and with a sufficient accuracy.
Until not long ago, such measurements were feasible only in the terahertz domain.
Nowadays, attosecond technology provides excellent means for measuring optical waveforms.
In the visible domain, attosecond streaking has recently reached subattosecond accuracy~\cite{Ossiander_2017_NP_13_280}.
For longer near-IR and mid-IR wavelengths, electro-optical sampling is an attractive alternative that does not require a vacuum setup~\cite{Keiber_2016_NP_10_159}.
A general name for the class of measurements enabling a direct access to the polarization response in the optical domain is \emph{attosecond polarization spectroscopy}.

\begin{figure}[!ht]
  \includegraphics[width=\linewidth]{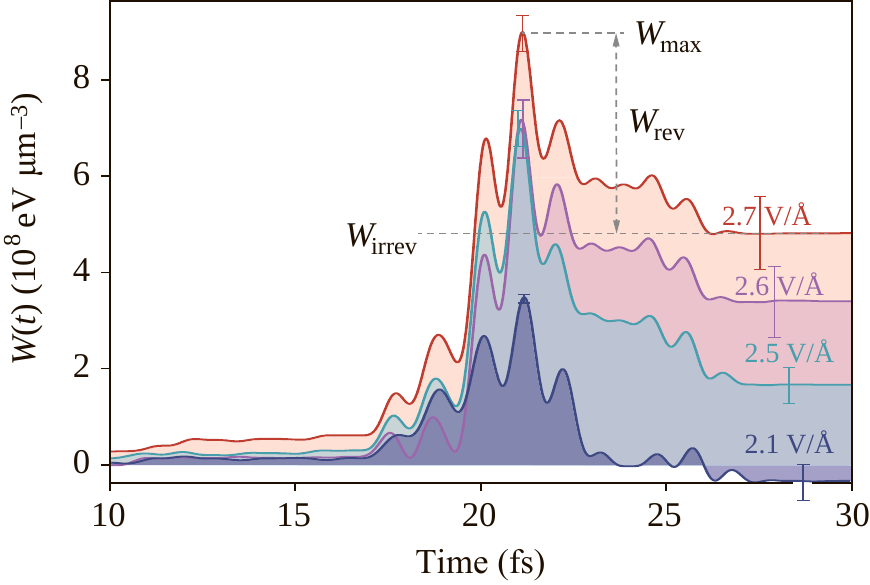}
  \caption{\label{f:Wt}%
    (Color online) Nonlinear energy exchange between a 3.1-fs 750-nm pulse and a fused-silica sample, measured by attosecond polarization spectroscopy.
    Adapted from~\cite{Sommer_2016_Nature_534_86}.
  }
\end{figure}

Figure~\ref{f:Wt} shows the real-time energy exchange between an intense few-cycle infrared laser pulse and fused silica, which was measured using attosecond polarization spectroscopy~\cite{Sommer_2016_Nature_534_86}.
To prepare this figure, the linear component of the polarization was subtracted from $\vec P(t)$, and the nonlinear component of the work performed by the electric field was evaluated by integrating Eq.~\eqref{e:dWdt}.
These results reveal that the amount of energy irreversibly dissipated in the sample crucially depends on the maximum applied field strength.
At $2.1$~V/\AA{}, the dissipated energy is smaller than the uncertainty of its nominal value, while the peak energy transiently transferred from the light pulse to the sample is comparable to that measured at the highest laser intensity, which was approximately 10\% below the damage threshold.
For peak laser fields stronger than $2.1$~V/\AA{}, a significant amount of energy is transferred from light to matter in the form of residual electronic excitations.
We denote this work, accumulated by the end of the interaction, as $W_\mathrm{irrev}$.
The difference between the peak value of the work, $W_\mathrm{max}$, and $W_\mathrm{irrev}$ characterizes the transient nonlinear energy transfer.
At this moment, it is an open question how to exploit a transient transfer of a large amount of energy from light to a medium for metrology or signal processing, but we recognize such reversible processes as an opportunity for future applications.

\subsection{Adiabatic evolution and topology of Bloch bands}\label{s:Berry}

Geometric and topological properties of Bloch bands, which originate from the Berry connection $\vec\xi_{nn}(\vec k)$, enrich the physics of electron motion steered by an electromagnetic field.
In particular, if the Berry curvature, defined by
\begin{equation*}
  \vec \Omega_n(\vec k) = \nabla\times \vec\xi_{nn}(\vec k),
\end{equation*}
is nonzero, then the velocity of a wavepacket accelerated by an electric field acquires an additional component known as the \emph{anomalous velocity}~\cite{Karplus_1954_PR_95_1154}:
\begin{equation*}
  \vec v_n^\mathrm{(a)} = - \dot{\vec K} \times \vec\Omega_n(\vec K) = \frac{e}{\hbar} \vec F(t) \times \vec\Omega_n(\vec K).
\end{equation*}
In the absence of the magnetic field, the anomalous velocity is orthogonal to the electric field.
In the presence of both electric $\vec F(t)$ and magnetic $\vec B(t)$ fields, the motion of an electron wavepacket obeys the following equations [cf. Eq.~\eqref{e:Bloch}]~\cite{Sundaram_1999_PRB_59_14915}:
\begin{subequations}\label{e:BlochBerry}
  \begin{align}
  \dot{\vec K} &= -\frac{e}{\hbar} [\vec F(t) + \dot{\vec r}_n\times \vec B(t)]
  ,\\
  \dot{\vec r}_n &= \frac{1}{\hbar} \nabla_{\vec K} E_n(\vec K) - \dot{\vec K} \times \vec\Omega_n(\vec K).
  \end{align}
\end{subequations}

The Berry connection also determines the geometric phase, which we defined in Eq.~\eqref{e:gammank}.
For a long time, it was argued that such a phase factor accumulated by a wavefunction during adiabatic evolution is physically meaningless and can be removed by a gauge transform.
In his seminal paper, \cite{Berry_1984_PRSA_392_45} showed that this is true only if the path followed by a set of adiabatic parameters $\vec\Lambda(t)$ remains open.
If the path is closed, i.e., returns to its starting point $\vec\Lambda(t_0)$, then the accumulated phase change is gauge invariant and therefore presents a physical observable.
It was then realized that the geometric phase $\gamma_{n, \vec k}(t, t_0)$ of an electron in a Bloch band is responsible for measurable changes of material properties, e.g., polarization~\cite{King-Smith_1993_PRB_47_1651, Resta_1994_RMP_66_899}, orbital magnetization~\cite{Thonhauser_2005_PRL_95_137205}, and density of states~\cite{Xiao_2005_PRL_95_137204}.
The concept of geometric phase also offered a new physical approach to such phenomena as the quantum Hall, Aharonov--Bohm, and Jahn--Teller effects~\cite{Bohm_2013}.
The integration of the geometric phase or the Berry curvature over the Brillouin zone yields topological invariants of Bloch bands (see Appendix~\ref{s:topology} for more details).

The role of the geometric phase in the interaction of a solid with a constant external electric field was studied by~\textcite{Zak_1989_PRL_62_2747}, who identified it in the Wannier--Stark ladder.
Without neglecting $\gamma_{n, \vec k}(t, t_0)$ in Eq.~\eqref{e:Houston}, the quantization of Bloch oscillations yields the following generalized expression for the Wannier--Stark ladder~\cite{Resta_2000_JP_12_R107, Bohm_2013, Lee_2015_PRB_92_195144}:
\begin{equation}\label{e:EnkFull}
  \mathcal E_{n,\ell} = \overline{E}_n + e a F_0 \left(\ell + \frac{\gamma_{n}^\mathrm{(Z)}}{2\pi}\right)
.
\end{equation}
Here,
\begin{equation*}
  \gamma_n^\mathrm{(Z)} = \oint_{\partial \Sigma} \vec \xi_{nn}(\vec k)\cdot d\vec k
\end{equation*}
is known as the Zak phase, and the integral is taken over an arbitrary closed loop $\partial\Sigma$ going around the entire Brillouin zone.

Unlike the Berry connection $\vec\xi_{nn}(\vec k)$, the Zak phase and the Berry curvature do not depend on the gauge (phase choice) of the Bloch amplitudes $u_{n, \vec k}(\vec r)$ and thus can be measured.
Experimental studies of phenomena related to a geometric phase form an active field of research in artificial periodic systems.
Recent works have demonstrated the measurements of the Zak phase~\cite{Atala_2013_NP_9_795, Duca_2015_Sci_347_6219} and oscillations of effective mass of a particle in an optical lattice driven by external force~\cite{Chang_2014_PRL_112_170404}.
In a recent paper~\cite{Li_2016_Science_352_1094}, Bloch state tomography was applied for measurements of the Berry curvature and topological invariants in the case of degenerate bands.

During the past decades, similar studies in solids were primarily focused on various kinds of the Hall effect, especially in 2D systems (quantum wells and graphene-like structures) exposed to strong magnetic fields~\cite{Klitzing_1980_PRL_45_494, Thouless_1982_PRL_49_405, Haldane_1988_PRL_61_2015, Nagaosa_2010_RMP_82_1539}.
The emerging possibilities to drive and track electron motion on time scales where momentum scattering is negligible enable studies of the anomalous group velocity and topological properties of Bloch bands with strong electric fields.
For example, effects of the Zak phase in graphene exposed to circularly polarized femtosecond pulses were measured using ARPES~\cite{Liu_2011_PRL_107_166803} and theoretically investigated by~\textcite{Kelardeh_2016_PRB_93_155434}.
The Berry curvature manifests itself in the electric current induced by the anomalous velocity component in a quantum well exposed to a circularly polarized ultrashort laser pulse~\cite{Virk_2011_PRL_107_120403}.
Similar ideas were recently realized in experiments on undoped GaAs quantum wells~\cite{Priyadarshi_2015_PRL_115_257401} and atomically thin MoS$_2$~\cite{Liu_2017_NP_13_262}.

\section{Summary and Outlook}\label{s:4}

The ultimate physical speed limits of electron-based metrology and signal processing are defined by how fast the electric or optical properties of materials can be manipulated.
Direct time-resolved access to the underlying phenomena in bulk systems and a comprehensive insight into their interaction with electromagnetic radiation on extremely short timescales are the keys to clarifying and, possibly, pushing these limits.
Attosecond techniques promise such access and insight.

An important opportunity for metrological applications, such as optical-field sampling, is the possibility to confine the excitation of charge carriers to a time interval shorter than a femtosecond.
Subsequent steering of carrier motion with the electric field of light enables subcycle trajectory control, which is the basis of attosecond metrology.
Quantum coherence plays an important role on a few-femtosecond time scale, even if experiments are performed at room temperature.
How to take advantage of the quantum nature of a light-driven electron wavepacket is one of the major open questions for future applications.

While excitation of charge carriers can be very fast, their interband recombination is a very slow process.
Especially for signal processing, phenomena that enable light-field-controlled manipulation of physical properties of a solid without substantial electronic excitations are going to play a central role.
Such phenomena imply reversible energy transfer, where the quantum states of electrons adapt to the external field, while the properties of ``field-dressed'' or adiabatic states significantly differ from those of the unperturbed ones.
This perspective is also the reason why we emphasize the concept of adiabaticity in this Colloquium.
From the experimental side, the recently emerged capability of measuring the energy flow between light and matter with attosecond resolution is likely to play an instrumental role in minimizing irreversible and maximizing transient nonlinear energy transfer.

The current focus of petahertz photonics is the exploration of relevant phenomena using the tools that have recently become available.
These phenomena include well-known effects, such as interband tunneling, and recently discovered ones, such as high-harmonic generation in solids.
Since dissipation and dephasing present major obstacles to potential applications, there is a strong demand for a better understanding of their microscopic origins.
Recent experiments indicate that the presence of a strong field may significantly accelerate the decay of interband coherence.
By exploring the underlying physics, future research should clarify the possibilities for coherent manipulation of electron wavepackets in solid-state materials.

We presented an up-to-date classification of laser--matter interaction regimes summarizing both classical results and recent discoveries in strong-field physics.
Even though our scheme may not exhaust all the possibilities, it serves as guidance for well-known strong-field phenomena and offers a framework for analyzing effects that may become important in the future.

Summarizing our outlook, a better understanding of how electrons behave in new regimes of light-matter interaction may open up new paths toward performing electronic operations with both unprecedented speed and low energy deposition into the material.
This may, in the long run, pave the way to petahertz electronic and photonic devices.
In the nearer future, these research efforts will spawn novel high-speed metrology based on solid-state detectors.

\appendix

\section{Ponderomotive energy in the tight-binding approximation}
\label{s:Up}

Breakdown of the EMA for carriers that approach the edges of BZ and do not tunnel to higher bands requires a generalized definition of the ponderomotive energy.
Substitution of Eq.~\eqref{e:En} into the general formula~\eqref{e:Up} yields the following analytical result:
\begin{equation}\label{e:UpTB}
  \Up = \sum_{\ell = 0}^{\ellmax} \varepsilon_{\mathrm{cv},\ell} J_0\left(\frac{\ell\w{B}}{\w0}\right) - \E{g},
\end{equation}
where $\varepsilon_{\mathrm{cv},\ell} = \varepsilon_{\mathrm{c},\ell} - \varepsilon_{\mathrm{v},\ell}$ is the difference of hopping integrals for conduction and valence bands.

\begin{figure}[!ht]
  \includegraphics[width=\linewidth]{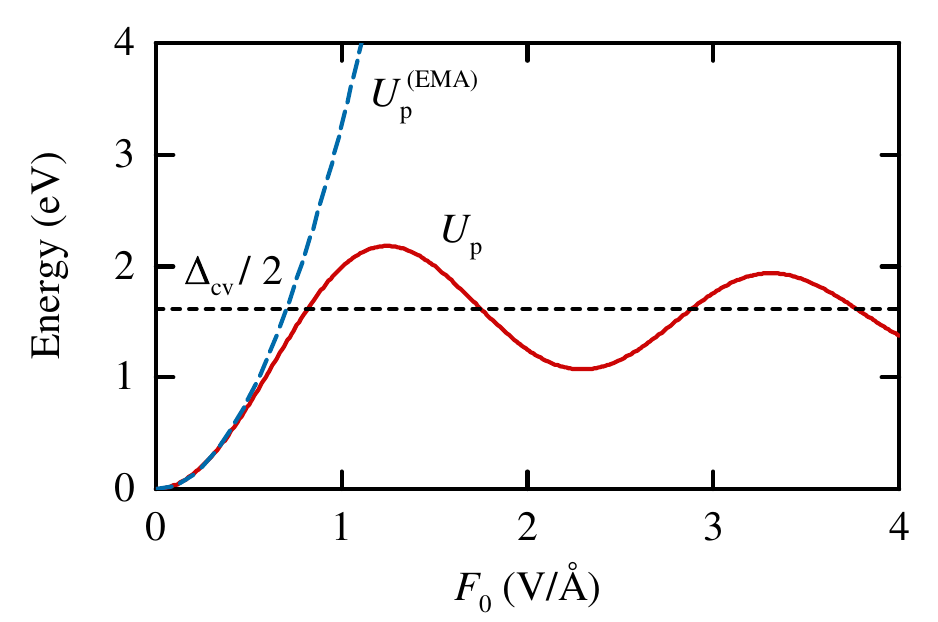}
  \caption{\label{f:Up_F0}%
    (Color online) Ponderomotive energy of an electron-hole pair moving along the $\Gamma$--M direction of SiO$_2$ in the lowest conduction and highest valence bands, calculated in the EMA (dashed curve), as well as for the exact band dispersion (solid curve).
    At high fields, where the electron approaches the BZ edges and stays in the same band, $\Up$ oscillates around one-half of the difference between the bandwidths of the conduction and valence bands $\Delta_\mathrm{cv}/2$ (short-dashed line).
  }
\end{figure}

As expected, in the weak-field limit, Eq.~\eqref{e:UpTB} follows a parabolic dependence on the field amplitude (Fig.~\ref{f:Up_F0}), while in the strong-field limit $\Up$ deviates from it and starts oscillating around a half difference of the conduction and valence band widths:
\begin{equation*}
\frac{\Delta_\mathrm{cv}}{2} = \frac{\Delta_\mathrm{c} - \Delta_\mathrm{v}}{2}
= \sum_{\ell = 1}^{\ellmax} \varepsilon_{\mathrm{cv},\ell}.
\end{equation*}

\section{Arbitrary-order corrections of adiabatic perturbation theory}
\label{s:Dyson}

Introducing the interband interaction potential $\widetilde V_\mathrm{cv}(t_n) = V_\mathrm{cv}(t_n) \e^{\ii \phi'_\mathrm{cv}(\vec k, t_n, t_0)}$ and the time-ordering operator $\hat{\mathcal T}$, one derives the $n$th order correction~\cite{Sakurai_2013}
\begin{multline}\label{e:ain}
  a_{\mathrm{c}, \vec k}^{(n)}(\vec k, t) = \left(-\frac{\ii}{\hbar}\right)^n
  \int_{t_0}^{t} \dd t_1 \ldots
\\
  \int_{t_0}^{t_{n-1}} \dd t_n\,
  \widetilde V_\mathrm{cv}(t_1) \ldots \widetilde V_\mathrm{cv}(t_n)
\end{multline}
and the time evolution operator
\begin{equation}
  \hat U(t, t_0) = \hat{\mathcal T}\exp\left[
   -\frac{\ii}{\hbar}
    \int_{t_0}^{t} \widetilde V_\mathrm{cv}(\tau)\,\dd\tau
  \right].
\end{equation}

\section{Hybrid Kane functions}
\label{s:Kane}

Wavefunctions of electrons in the lattice tilted by a static external field can be approximated using the \emph{hybrid Kane functions}~\cite{Kane_1960_JPCS_12_181, Fritsche_1966_PSSB_13_487, Glutsch_2004}
\begin{equation}\label{e:Kane}
  \varphi_{n,l}^\mathrm{(K)}(\vec k_\perp, \vec r) = \int_{-\pi/a}^{\pi/a}
    \eta_{n, l}(\vec k) \psi_{n,\vec k}^\mathrm{(B)}(\vec r)\,\dd k_x,
\end{equation}
where the transformation of Bloch functions $\psi_{n,\vec k}^\mathrm{(B)}(\vec r) = u_{n,\vec k}(\vec r) \e^{\ii\vec k\cdot\vec r}$ to real space is taken only over the $k_x$ component, which is directed along the field polarization.
Here,
\begin{equation}\label{e:eta}
  \eta_{n, l}(\vec k) = \exp\left\{ -\frac{\ii}{ e F_0}
  \int_{0}^{k_x} \left[ \mathcal E_{n,l}(\vec k_\perp) - E'_n(\vec k) \right] \dd k_x \right\}
\end{equation}
are the solutions of an adiabatic eigenvalue problem
\begin{equation}\label{e:KaneEq}
  \left[E'_n(\vec k) + \ii e F_0 \partial_{k_x} \right] \eta_{n,l}(\vec k)
= \mathcal E_{n,l}(\vec k_\perp) \eta_{n, l}(\vec k).
\end{equation}
for the Hamiltonian
\begin{equation}
  \hat H = -\frac{\hbar^2 \nabla^2}{2m_0} + \hat V_\mathrm{latt}(\vec r) + eF_0 x
\end{equation}
in the momentum representation and in the single-band approximation $x = \partial_{k_x} + \ii X_{nn}(\vec k)$, $X_{nm}(\vec k) = 0, n \ne m$.

The continuous part of $\mathcal E_{n,l}(\vec k_\perp)$ [see Eq.~\eqref{e:KaneE}] is given by
\begin{equation}\label{e:oEn}
  \overline{E}'_n(\vec k_\perp) = \frac{a}{2\pi} \int_{-\pi/a}^{\pi/a} E'_n\brK{t}\,\dd k_x,
\end{equation}
where
\begin{align*}
  E'_n(\vec k) &= E_n(\vec k) + eF_0 X_{nn}\brK{t}
,\\
  \vec K(t) &= K_x(t) + \vec k_\perp
,\quad
  K_x(t) = \left(k_x - \frac{e F_0}{\hbar} t\right)\ \mathrm{mod}\ G
,\\
  X_{nn}(\vec k) &= \langle u_{n,\vec k} | \ii\partial_{k_x} | u_{n,\vec k}\rangle .
\end{align*}

The Kane functions are orthonormal and complete:
\begin{equation}
  \int_{-\infty}^{+\infty} \varphi_{n,l}^\mathrm{(K)*}(\vec k_\perp, \vec r)
  \varphi_{n',l'}^\mathrm{(K)}(\vec k_\perp, \vec r)\,\dd^3 r = \delta_{nn'} \delta_{ll'}
,
\end{equation}
and they satisfy the ladder property:
\begin{equation}
  \varphi_{n,l}^\mathrm{(K)}(\vec k_\perp, \vec r) = \varphi_{n,0}^\mathrm{(K)}(\vec k_\perp, \vec r - l a).
\end{equation}
The hybrid Kane functions converge to the Wannier functions
\begin{equation}
  w_{n, l}(\vec k_\perp, \vec r) = \frac{a}{2\pi} \int_{-\pi/a}^{\pi/a}
\psi_{n,\vec k}^\mathrm{(B)}(\vec r) \e^{-\ii k_x la}\,\dd k_x
\end{equation}
in the limit of the infinite electric field
\begin{equation}
  \lim_{F_0 \rightarrow \infty} \varphi_{n,l}^\mathrm{(K)}(\vec k_\perp, \vec r) = w_{n,l}(\vec k_\perp, \vec r)
\end{equation}
or in the limit of dispersionless bands $E_n(\vec k) = \mathrm{const}$, which is true for the core states~\cite{Glutsch_2004}.

\section{Berry curvature and topological invariants}
\label{s:topology}

In general, the Berry curvature is not a vector, but a rank-2 antisymmetric tensor defined as an exterior derivative of the Berry connection
\begin{equation}\label{e:Omn}
  \Omega_{\mu\nu} = \partial_{\mu} \xi_{\nu} - \partial_{\nu} \xi_{\mu} + \ii [\xi_\mu, \xi_\nu],
\end{equation}
similarly to the electromagnetic field tensor~\cite{Landau_1975}.
Here, the band indices were omitted for simplicity, $\nu$ and $\mu$ enumerate coordinates, and $\partial_\mu \equiv \partial/\partial k_\mu$.

Unlike the electromagnetic field, which has an Abelian gauge symmetry $U(1)$, the Berry connection and curvature have multiple components in a subspace of degenerate Bloch bands, so that they become matrices with a non-Abelian gauge structure~\cite{Wilczek_1984_PRL_52_2111, Yu_2011_PRB_84_075119, Alexandradinata_2016_PRB_93_205104, Xiao_2010_RMP_82_1959, Li_2016_Science_352_1094}.
This results in a nonvanishing commutator between components of the Berry connection [see Eq.~\eqref{e:Omn}].

The Berry curvature is proportional to an imaginary part of a more general object known as the \emph{quantum geometric tensor}
\begin{equation}
  Q_{\mu\nu} = F_{\mu\nu} - \frac{\ii}{2} \Omega_{\mu\nu},
\end{equation}
which was initially introduced by~\textcite{Provost_1980_CMP_76_289} in the framework of a geometric approach to quantum mechanics.
The real part of this tensor defines the Fubini--Study metric $F_{\mu\nu}$, which characterizes the quantum distance between two infinitesimally separated states:
\begin{equation}
 \dd s^2 = 1 - |\langle \Psi(\vec k) | \Psi(\vec k + \dd \vec k) \rangle|^2 = \sum_{\mu\nu} F_{\mu\nu}
 \dd k^\mu \dd k^\nu.
\end{equation}

Modern developments of condensed matter theory employing the mathematical methods of differential geometry and topology have led to the formulation of the \emph{topological band theory}~\cite{Hasan_2010_RMP_82_3045, Bansil_2016_RMP_88_021004}, where Bloch bands are characterized with a new class of quantum numbers, the topological invariants~\cite{Nakahara_2003}.

An illustrative example of a topological invariant is the first Chern number $C_n$ counting the number of vortices (Dirac points) in the band.
It is defined by integration of the Berry curvature over a closed surface $\Sigma$ around the Brillouin zone~\cite{Gradhand_2012_JPCM_24_213202}
\begin{equation}
  C_n = \frac{1}{2\pi} \iint_\Sigma \vec\Omega_n(\vec k)\,\dd\vec S,
\end{equation}
where $\dd\vec S$ is an oriented surface element in reciprocal space.
If the Bloch functions are smooth with respect to $\vec k$, one can employ the Stokes theorem~\cite{Nakahara_2003} and express the Chern number via the Zak phase:
\begin{equation}\label{e:Cn}
  C_n = \frac{\gamma_n^\mathrm{(Z)}}{2\pi}.
\end{equation}

The Chern number appears in condensed matter physics as a parameter determining the quantization of Hall conductivity in a 2D electron gas~\cite{Thouless_1982_PRL_49_405, Xiao_2010_RMP_82_1959}
\begin{equation*}
  \sigma_{xy} = \frac{e^2}{\hbar} \sum_{n \in \mathrm{occ.}} C_n,
\end{equation*}
where the summation is taken over occupied bands.

From Eqs.~\eqref{e:EnkFull} and~\eqref{e:Cn}, it is clear that the Chern number emerges in the quantization law for the Wannier--Stark resonances~\eqref{e:EnkFull} and provides a classification between the topologically trivial ($C_n = 0$) and nontrivial ($C_n \ne 0$) cases~\cite{Lee_2015_PRB_92_195144}.
Recently, it has been proven~\cite{Brouder_2007_PRL_98_046402, Panati_2013_CMP_322_835} that the equality $C_n = 0$ for all bands gives a necessary and sufficient condition for the existence of the maximally localized Wannier functions~\cite{Marzari_2012_RMP_84_1419}.
For example, this is the case for systems with time-reversal symmetry.

The $\mathbb{Z}_2$ topological invariant $\nu$ encodes the time-reversal invariance properties of the bulk band structure~\cite{Kane_2005_PRL_95_146802, Fu_2006_PRB_74_195312} and distinguishes between the trivial ($\nu = 0$) and topological materials ($\nu = 1$).
In a 2D system, it can be expressed in the following general form~\cite{Soluyanov_2011_PRB_83_035108, Bansil_2016_RMP_88_021004}
\begin{equation}
  \nu = \frac{1}{2\pi} \sum_{n \in\mathrm{occ.}} \left[\oint_{\partial \tau} \vec\xi_{nn}(\vec k)\,\dd\vec k - \iint_\tau \vec\Omega_n(\vec k)\,\dd \vec S \right] \mathrm{mod}\,2,
\end{equation}
where the integrals are taken over one-half of the BZ $\tau$ and the boundary of that half $\partial\tau$.

\begin{acknowledgments}
This work was supported by the DFG Cluster of Excellence: Munich-Centre for Advanced Photonics (MAP).
S.\,Yu.\,K. acknowledges support from the Austrian Science Fund (FWF) within the Lise Meitner Project No. M2198-N30.

We thankfully acknowledge our colleagues, Dr. Eleftherios Goulielmakis and Dr. Nicholas Karpowicz, for valuable and stimulating discussions.

\end{acknowledgments}

\section*{Glossary of Abbreviations}

\renewcommand{\descriptionlabel}[1]{\hspace\labelsep #1}
\begin{description}[align=left,labelwidth=1.5cm]
\item[ARPES] Angle-resolved photoemission spectroscopy
\item[BH] Band hybridization (parameter)
\item[BP] Bloch-to-ponderomotive (ratio of energies)
\item[BZ] Brillouin zone
\item[CB] Conduction band
\item[CEP] Carrier-envelope phase
\item[CWRF] Carrier-wave Rabi flopping
\item[DFKE] Dynamic Franz--Keldysh effect
\item[DL] Dynamic localization
\item[EMA] Effective-mass approximation
\item[ERF] Envelope Rabi flopping
\item[FWHM] Full-width at half-maximum
\item[IR] Infrared radiation
\item[HHG] High-order harmonic generation
\item[MOSFET] Metal-oxide-semiconductor field-effect transistor
\item[NP] Nonperturbative
\item[PHz] Petahertz (frequency)
\item[RB] Rabi-to-Bloch (ratio of frequencies)
\item[RF] Rabi flopping
\item[RWA] Rotating-wave approximation
\item[SC] Semiclassical
\item[SFA] Strong-field approximation
\item[TDSE] Time-dependent Schr\"odinger equation
\item[THz] Terahertz (frequency)
\item[VIS] Visible (radiation)
\item[WS] Wannier--Stark (ladder, resonances, states)
\end{description}

\bibliographystyle{apsrmp4-1}
\bibliography{Kruchinin_etal_RMP}

\end{document}